\documentclass[11pt]{article}
\usepackage{amsmath,amssymb}
\usepackage{natbib}
\usepackage[margin=1in]{geometry}
\usepackage{graphicx}
\usepackage[ruled,vlined,linesnumbered]{algorithm2e}

\usepackage{subcaption}
\usepackage[dvipsnames]{xcolor}
\usepackage{colortbl}
\usepackage{enumitem}
\usepackage[normalem]{ulem}

\usepackage{multirow}
\usepackage{bm}
\usepackage{hyperref}
\usepackage{url}
\newcommand{\vect}[1]{\bm{#1}}
\newcommand{\matr}[1]{\bm{\mathsf{#1}}}

\SetAlgoHangIndent{1.5em}  
\SetInd{0.5em}{1.5em}      

\title{Transferable inference of turbulence models for urban flows with the Parameter-Regularised Ensemble Kalman Filter}

\author{
Emanuele Bombardi$^{1,2}$, Andrea Nóvoa$^{3,4}$, Luca Magri$^{3,5,6}$, Alessandro Parente$^{1,2,7}$\\[1ex]
\small $^1$Université Libre de Bruxelles, Aero-Thermo-Mechanics Laboratory, Brussels, Belgium\\
\small $^2$ULB \& VUB, Brussels Institute for Thermal-Fluid Systems and Clean Energy (BRITE), Brussels, Belgium\\
\small $^3$Department of Aeronautics, Imperial College London, London, United Kingdom\\
\small $^4$I-X, Imperial College London, White City, London, United Kingdom\\
\small $^5$DIMEAS, Politecnico di Torino, Torino, Italy\\
\small $^6$Brussels Institute for Advanced Studies (BrIAS), Brussels, Belgium\\
\small $^7$WEL Research Institute, Wavre, Belgium\\[1ex]
\small Corresponding author: Luca Magri, \texttt{l.magri@imperial.ac.uk}
}
\date{}

\begin{document}
\maketitle

\begin{abstract}
The accurate simulation of urban flow is key to designing building ventilation, understanding cities' micrometeorology, and predicting pollutant dispersion. Reynolds-Averaged Navier–Stokes (RANS) simulations are a common modelling approach for simulating urban flow, but their accuracy depends on the closure model and its parameters. These parameters are inferred from benchmark cases, but they are not necessarily suitable for realistic urban environments, which involve different physical mechanisms. This is referred to as {\it the transferability} problem of RANS urban modelling. The objective of this work is to propose a robust Bayesian method to {sequentially} infer  RANS parameters  for urban flow modelling. Key to the approach is the mathematical derivation of the parameter-regularised ensemble Kalman filter (PR-EnKF), which is the analytical solution of the data assimilation problem for the sequential parameter estimation. The cost functional is regularised using the prior knowledge on the turbulence parameters, thereby ensuring that the Bayesian updates remain   within physical ranges. The parameters are first inferred on an isolated building, and then transferred to three cases of increasing complexity: (i) a high-rise building, (ii) a multi-building array, and (iii) the Shinjuku district urban environment~\citep{AIJ_CFD_Guide_2007}.  Results show that the PR-EnKF achieves faster convergence, reducing parameter uncertainty by an order of magnitude and reconstruction errors by up to 50\%. {Because of the regularisation, the PR-EnKF selectively updates the most important parameters.} This work enables robust large-scale urban flow simulation whilst reducing the computational overhead of model optimisation for urban planning and air quality assessment.
\end{abstract}


\section{Introduction}
\label{sec:introduction}

Sustainable urban design requires an understanding of atmospheric flow dynamics, including wind patterns, pollutant transport, and thermal conditions. Reynolds-Averaged Navier-Stokes (RANS) computational fluid dynamics simulations are an established tool for predicting such complex flow phenomena in urban engineering applications \citep{Blocken2015, Kumar2024, Bombardi2025}. Although Large Eddy Simulation provides high-fidelity resolution of turbulent structures, and GPU-accelerated Lattice Boltzmann methods are emerging as a viable high-accuracy alternative \citep{Latt2021, Hu2026}, two-equation RANS closures remain the practical choice for large-scale urban applications \citep{Tominaga2009, Blocken2018}. 
Notwithstanding, RANS predictive accuracy is limited by the assumptions of turbulence models and uncertainties in the closure parameters, in which empirical coefficients should capture  turbulent transport physics  \citep{BayesianUQ_RANSChannelMoser2011, Edeling2014, Xiao2015BayesianRANS, Iaccarino2017}. These parameters arise from tensor invariance and dimensional consistency,  but their values are empirically calibrated against canonical flows \citep{Pope_2000}. In complex geometries such as urban environments, these standard {parameters} require adjustments to achieve quantitative accuracy \citep{TOMINAGA20081749}.

\subsection{Parameter optimisation and inference in urban flows}
\label{subsec:paradox}
Accurate predictions are critical for large urban domains, but parameter optimisation on geometries with tens of millions of cells remains computationally prohibitive. {Therefore,  parameters are optimised (i.e., inferred) on simplified geometries and then deployed (i.e., \emph{transferred}) to realistic city-scale simulations. This means that the optimised parameters must be suitable even when applied to geometries that differ from those used in optimisation.} We refer to this property as \emph{parameter transferability}, which represents the core challenge in urban flow modelling. Bridging this gap between computationally feasible optimisation cases and complex urban applications demands robust parameter estimation frameworks that ensure cross-geometry {transferability}.

Traditional optimisation approaches often produce parameter values that deviate significantly from physically meaningful ranges. Reported deviations span multiple orders of magnitude: fivefold increases for wind turbine flows \citep{Rocha2013}, threefold variations in atmospheric boundary layer simulations \citep{ref51}, and two- to thirteen-fold deviations in urban applications \citep{ref56, SHIRZADI2020101756, VILLANUEVA2024}. These geometry-specific optimisations achieve high accuracy on the optimisation case but do not necessarily transfer to  realistic cases:  \citet{Zhao2022} demonstrated substantial performance degradation when transferring parameters from single-building optimisations to building arrays. This limited transferability arises from the non-convex optimisation landscape, which is  characterised by multiple local minima {in the parameter inference problem}. Without explicit constraints, optimisation algorithms converge to geometry-specific parameters that lack physical consistency across different flow configurations.\\

{Data assimilation techniques \citep[e.g.,][]{Daley1991AtmosphericData} offer a principled framework for combining model predictions with observations while accounting for uncertainties. Recent applications have demonstrated the effectiveness of both variational \citep{Zaki2025, WangZaki2025} and ensemble-based approaches for state estimation \citep{Meldi2017}, parameter inference \citep{VILLANUEVA2024}, and flow reconstruction \citep{LeProvost_Eldredge}. Among ensemble methods, the Ensemble Kalman Filter (EnKF) \citep{Kalman, Anderson2001, Evensen2009}, which performs a joint state-parameter update via an augmented state vector,  has become a useful tool for parameter estimation in fluid problems. Variants include ES-MDA \citep{emerick2012, emerick2013ensemble}, which iterates the ensemble update over a fixed observation window, and the Dual EnKF \citep{Moldovan2022}, which alternates separate Kalman updates for the flow state and model parameters within each assimilation cycle. 

Ensemble methods rely on Gaussian assumptions and linear updates that, in the absence of explicit regularisation, allow parameter estimates to change arbitrarily from physically meaningful ranges, particularly with sparse or noisy observations \citep{GORLE2015202, Stroud, Novoa2022}. Various approaches address these limitations, including post-update corrections and parameter transformations \citep{e23121673, Iglesias_2016}, but they often lack a theoretical foundation and may not preserve Bayesian consistency, compromising the probabilistic coherence between prior information, observations, and posterior estimates that ensures optimal inference. Recent advances in physics-informed machine learning \citep{YANG2021109913, WU2024124678} and comprehensive reviews of turbulence modelling \citep{Vinuesa2022, Xiao2019} emphasise that incorporating physical constraints into Bayesian frameworks \citep{Jiang2021, Cherroud2024} can provide an effective regularisation to prevent overfitting whilst preserving physical consistency. In this regard, penalty terms have successfully been applied within ensemble Kalman frameworks to enforce smoothness and stability constraints \citep{ZHANG2020109517,ZhangJFM2024}{, to infer model errors with the Bias-Aware Ensemble Kalman filter \citep{NOVOA2024116502},} or to constrain neural network predictions of model coefficients \citep{LiuJFM2026}. In contrast, the present work directly regularises the  turbulence model parameters with the objective of promoting cross-geometry transferability. \\}

This work develops a Parameter-Regularised Ensemble Kalman Filter (PR-EnKF) that incorporates prior turbulence modelling knowledge about model parameters directly into the Kalman filtering process. We derive a modified EnKF formulation with an additional regularisation term in the cost function, penalising deviations of parameters from their literature-consistent values. {These values, taken from prior calibration and accepted practice, act as a prior that imposes a soft constraint on the inference, keeping the estimated parameters within  meaningful ranges.}
The methodology is validated through an {assimilation}-transferability study on four urban-flow configurations of increasing complexity: parameters are {inferred} on the CEDVAL A1-1 isolated building \citep{CedvalHam}, then transferred to the AIJ high-rise building, AIJ building array, and the full-scale Shinjuku district in Tokyo \citep{AIJ_CFD_Guide_2007}. \\

The remainder of this paper is organised as follows. Section~\ref{sec:framework} presents the Bayesian  framework for the PR-EnKF, including the analytical solution. Section~\ref{sec:validation} discusses implementation details,  computational aspects, the CFD configuration and test cases. Section~\ref{sec:results} presents results and analyses on canonical test cases. Section \ref{sec:results_transfer} investigates a realistic large-scale urban flow.  Section~\ref{sec:conclusions} concludes the paper. 

\section{Mathematical framework}
\label{sec:framework}

This section presents the mathematical derivation of the Parameter-Regularised Ensemble Kalman Filter (PR-EnKF). We set up the problem in the state-space formulation  (\S\ref{subsec:problem}), review the standard Ensemble Kalman Filter (\S\ref{subsec:standard_enkf}), and derive the parameter-regularised filter  (\S\ref{subsec:regularised_enkf}).

\subsection{Problem formulation}
\label{subsec:problem}

{
We consider a set of deterministic steady equations formulated as an iterative nonlinear state-space model
\begin{equation}
\vect{\phi}^{n+1} = \mathcal{F}\left(\vect{\phi}^{n}, \vect{\alpha}^{n}\right) 
\label{eq:state_space_steady}
\end{equation}
where the superscript 
$n$ denotes the iteration index; 
 $\mathcal{F}: \mathbb{R}^{N_\phi} \rightarrow \mathbb{R}^{N_\phi}$ encapsulates the governing equations and boundary conditions; 
and $\vect{\phi}^{n}\in \mathbb{R}^{N_\phi}$ and $\vect{\alpha}^{n}\in \mathbb{R}^{N_\alpha}$ are the state vector and model parameters at iteration $n$, respectively.  
In this paper, \eqref{eq:state_space_steady} describes the iterative progression of the RANS solver toward a steady-state solution. The state $\vect{\phi}$ comprises the discretised flow variables (e.g., velocity components, pressure, and turbulence quantities) at all grid points, while  $ \vect{\alpha} $ contains turbulence model parameters (e.g., eddy-viscosity constants). The operator $\mathcal{F}$ represents one nonlinear solver iteration (e.g., using SIMPLE or PISO-based schemes), including spatial discretisation, linearisation, and solution of the resulting algebraic system.  \\

The aim of this work is to integrate observations into the iterative RANS solver via data assimilation to 
(i) accelerate convergence toward the steady state, and 
(ii) {infer} the parameters $\vect{\alpha}$ to ensure consistency with experimental data and enhance transferability across environments. 
To bridge the gap between the deterministic RANS operator and the uncertainties inherent in physical modelling \citep{kennedy2001bayesian}, we cast the assimilation problem in a stochastic framework. 
First, we model aleatoric uncertainties in the state and parameters with stochastic errors, so \eqref{eq:state_space_steady} becomes
\begin{equation}
\vect{\phi}^{n+1} = \mathcal{F}\left(\vect{\phi}^{n} + \vect{\epsilon}_\phi, \vect{\alpha}^{n} + \vect{\epsilon}_\alpha\right) 
\label{eq:stochastic_model}
\end{equation}
where the error terms are modelled as normal distributions with zero means, i.e., 
$
\vect{\epsilon}_\phi \sim \mathcal{N}(\vect{0}, \matr{C}_{\phi\phi})
$
and 
$\vect{\epsilon}_\alpha \sim \mathcal{N}(\vect{0}, \matr{C}_{\alpha\alpha})
$. 
Second, because  the state vector $\vect{\phi}$ is high-dimensional, we observe the system at a limited number of spatial locations $N_q \ll N_\phi$. The sparse observations $\vect{q} \in \mathbb{R}^{N_q}$ are linked to the full state via a measurement operator $\mathcal{M}: \mathbb{R}^{N_\phi} \rightarrow \mathbb{R}^{N_q}$
\begin{equation}
\vect{q}^{n} = \mathcal{M}(\vect{x}, \vect{\phi}^{n}) + \vect{\epsilon}_q
\label{eq:measurements}
\end{equation}
where $\vect{\epsilon}_q \sim \mathcal{N}(\vect{0}, \matr{C}_{qq})$ accounts for measurement or representation errors (e.g., spatial discretisation or sensor limitations). 
Third, we define the augmented state vector $\vect{\psi} = [\vect{\phi}; \vect{\alpha}; \vect{q}]$, which comprises the state variables, the model parameters, and the model observables. Thus, the problem can be reformulated as 
\begin{equation}
\begin{cases}
 \vect{\phi}^{n+1} = \mathcal{F}(\vect{\phi}^{n}+ \vect{\epsilon}_\phi, \vect{\alpha}^{n} + \vect{\epsilon}_\alpha) \\
 \vect{\alpha}^{n+1}  =  \vect{\alpha}^{n} \\
 \vect{q}^{n+1} =  \vect{q}^{n} 
\end{cases}
 \Leftrightarrow \quad \vect{\psi}^{n+1} = \mathbf{F}(\vect{\psi}^n+ \vect{\epsilon}_\psi), 
 \label{eq:augmented_state}
\end{equation}
where $\mathbf{F}$ and $\vect{\epsilon}_\psi$ are the augmented nonlinear operator and aleatoric uncertainties, respectively. Formulation \eqref{eq:augmented_state}  allows for the definition of the following linear operators
\begin{equation}
\matr{M}_q = \begin{bmatrix} \matr{0}_{(N_\phi+N_\alpha)\times N_q} & \matr{I}_{N_q} \end{bmatrix}, \quad \mbox{such that} \quad \matr{M}_q\vect{\psi} = \vect{q},
\label{eq:measurement_operator}
\end{equation}
\begin{equation}
\matr{M}_\alpha = \begin{bmatrix} \matr{0}_{N_\phi\times N_q} & \matr{I}_{N_\alpha} & \matr{0}_{N_q\times N_q} \end{bmatrix}, \quad \mbox{such that} \quad \matr{M}_\alpha\vect{\psi} = \vect{\alpha},
\label{eq:parameter_operator}
\end{equation}
where $\matr{0}_{s_1\times s_2}$ is an $s_1 \times s_2$ matrix of zeros and $\matr{I}_{s}$ is the identity matrix of size $s$.\\

Data assimilation makes qualitatively accurate models quantitatively accurate by combining model predictions with observational data \citep[e.g.,][]{Magri2020, Nóvoa_Magri_2022}. Specifically, sequential data assimilation methods are designed to reduce the uncertainty in the state, parameters and observables by sequentially assimilating data in time. In this steady-state framework, the observation vector, $\vect{d} \in \mathbb{R}^{N_q}$, is constant across all iterations, and the assimilation is performed at intervals during the iterative process.
%
At each assimilation step $n_a$, the model state and parameters are updated by combining measurement data $\vect{d}$ with model estimates. The result is the analysis state, which  provides a statistically optimal estimate of the `true' state. 
We perform the data assimilation every $\Delta n_a>1$ iterations to allow the solver to adapt to the analysis parameters before incorporating the next correction.  From now on,  we drop the iteration superscript $n$ unless necessary for clarity.
}

\subsection{Ensemble Kalman filter}
\label{subsec:standard_enkf}
 From a Bayesian perspective, we seek the posterior distribution of the augmented state vector $\vect{\psi}$ conditioned on observations $\vect{d}$
 \begin{equation}
p(\vect{\psi}|\vect{d}) \propto p(\vect{d}|\vect{\psi}) \, p(\vect{\psi}),
\label{eq:bayes_theorem}
\end{equation}
where $p(\vect{d}|\vect{\psi})$ is the likelihood and $p(\vect{\psi})$ is the prior distribution. Under Gaussian assumptions, maximising the posterior is equivalent to minimising the cost function
\begin{equation}
\mathcal{J}(\vect{\psi}) = \left\|\vect{\psi}^f - \vect{\psi}\right\|_{\matr{C}_{\psi\psi}^{f^{-1}}}^2 + \left\|\vect{d} - \matr{M}_q\vect{\psi}\right\|_{\matr{C}_{d\!d}^{-1}}^2,
\label{eq:cost_function_standard}
\end{equation}
where the superscript $f$ stands for forecast {(here, referring to the model prediction at the current solver iteration rather than a temporal forecast)}, 
$\matr{C}_{dd}$ is the observation error covariance matrix, 
and $\|\cdot\|_{\matr{C}^{-1}}^2$ denotes the squared Mahalanobis distance with precision matrix $\matr{C}^{-1}$. 

{
To address the nonlinearity of the RANS solver and the high dimensionality of the state space, the distributions are approximated {with} a Monte Carlo ensemble of $N_e$ model realisations, so that the first two moments of $\vect{\psi}$ are estimated with the ensemble mean and covariance 
\begin{subequations}\label{eq:ens_stats}
\begin{align}
 \label{eq:ensemble_avg}
\mathbb{E}(\vect{\psi})& \approx \bar{\vect{\psi}}= \frac{1}{N_e}\sum_{m=1}^{N_e}
\vect{\psi}_m  \\ 
\matr{C}_{\psi\psi} = \begin{bmatrix}
    \matr{C}_{\phi\phi}   & \matr{C}_{\phi\alpha} & \matr{C}_{\phi q}\\
    \matr{C}_{\alpha\phi} & \matr{C}_{\alpha\alpha} & \matr{C}_{\alpha q} \\
    \matr{C}_{q\phi}     & \matr{C}_{q\alpha}   & \matr{C}_{qq}
\end{bmatrix} &\approx \frac{1}{N_e - 1}\sum_{m=1}^{N_e}
(\vect{\psi}_m - \bar{\vect{\psi}})
(\vect{\psi}_m - \bar{\vect{\psi}})^\top.
\label{eq:ensemble_cov}
\end{align}
\end{subequations}
With this, the minimisation of \eqref{eq:cost_function_standard} can be solved for each ensemble member independently, yielding  an  update equation for each member $m = 1,\ldots,N_e$
\begin{equation}
\vect{\psi}_m^a = \vect{\psi}_m^f + \matr{K}(\vect{d}_m - \matr{M}_q\vect{\psi}_m^f),
\label{eq:enkf_update}
\end{equation}
where $\matr{K} = \matr{C}_{\psi\psi}^f\matr{M}_q^\top(\matr{C}_{d\!d} + \matr{M}_q\matr{C}_{\psi\psi}^f\matr{M}_q^\top)^{-1}$ is the Kalman gain matrix  \citep{Evensen2009}. 
Each ensemble member $m$ is updated with a different $\vect{d}_m\sim\mathcal{N}(\vect{d}, \matr{C}_{dd})$ to avoid covariance underestimation \citep{burgers1998analysis}. 
The state and parameter update equations are
\begin{subequations}
\begin{equation}
\begin{cases}
\vect{\phi}_m^a = \vect{\phi}_m^f + {\matr{K}}_{\phi,d}(\vect{d}_m - \vect{q}_m^f) \\[.8em]
\vect{\alpha}_m^a = \vect{\alpha}_m^f + {\matr{K}}_{\alpha,d} (\vect{d}_m - \vect{q}_m^f)
\end{cases}
\label{eq:enkf_state_param_update}
\end{equation}
where the {matrices}
\begin{equation}
{\matr{K}}_{\phi,d} = \matr{C}_{\phi q}^f
        \left(\matr{C}_{d\!d}+\matr{C}_{qq}^f\right)^{-1} \quad\text{and }\quad 
{\matr{K}}_{\alpha,d} = \matr{C}_{\alpha q}^f\left(\matr{C}_{d\!d} + \matr{C}_{qq}^f\right)^{-1}
    \label{eq:gains_enkf}
\end{equation}
\end{subequations}
correct the state variables $\vect{\phi}$ and parameters $\vect{\alpha}$, respectively, using  flow observation innovations
$(\vect{d}_m - \vect{q}_m^f)$, according to the cross-covariances $\matr{C}_{\phi q}^f$ and $\matr{C}_{\alpha q}^f$. 
}

\subsection{Parameter-Regularised EnKF}
\label{subsec:regularised_enkf}
Standard ensemble Kalman filters can yield analyses that {are unphysical} or lack interpretability \citep[e.g.,][]{li2019constrained}. 
\citet{ZHANG2020109517} demonstrated that  regularising the data assimilation cost function can significantly stabilise the inversion and mitigate overfitting. 
Building on this principle, we introduce a literature-consistent parameter regularisation term into the EnKF---the Parameter-Regularised EnKF (PR-EnKF)--which discourages excessive changes of the parameters from established turbulence model coefficients. %
From a Bayesian perspective, we extend \eqref{eq:bayes_theorem} by incorporating literature-consistent parameter values $\vect{p}$
\begin{equation}
p(\vect{\psi}|\vect{d}, \vect{p}) \propto p(\vect{d}|\vect{\psi}) \, p(\vect{p}|\vect{\psi}) \, p(\vect{\psi}),
\label{eq:bayes_regularised}
\end{equation}
{
where $p(\vect{p}|\vect{\psi})$ is the likelihood of the literature-consistent parameters $\vect{p}$ given the augmented state $\vect{\psi}$, and we assumed statistical independence between $\vect{d}$ and $\vect{p}$, i.e., $p(\vect{d},\vect{p}) = p(\vect{d})p(\vect{p})$.  In practice, the literature-consistent parameters act as pseudo-observations, {i.e.,  values that are not measured in the flow but that are inherited from the literature}. 
We assume that the parameter likelihood is Gaussian distributed as 
$p(\vect{p}|\vect{\psi}) = \mathcal{N}(\matr{M}_\alpha\vect{\psi}, \matr{C}_{p\!p})$, where 
$\matr{C}_{p\!p} \in \mathbb{R}^{N_\alpha \times N_\alpha}$ is the pseudo-observation error covariance matrix, i.e.,  the uncertainty associated with  $\vect{p}$. 
Taking $-2\log(\cdot)$ of \eqref{eq:bayes_regularised} yields the parameter-regularised cost function
\begin{equation}
\mathcal{J}(\vect{\psi}) = \underbrace{\left\|\vect{\psi}^f - \vect{\psi}\right\|_{\matr{C}_{\psi\psi}^{f^{-1}}}^2}_{\text{Term 1}} + 
\underbrace{\left\|\vect{d} - \matr{M}_q\vect{\psi}\right\|_{\matr{C}_{d\!d}^{-1}}^2}_{\text{Term 2}} + 
\underbrace{\left\|\vect{p} - \matr{M}_\alpha\vect{\psi}\right\|_{\matr{C}_{p\!p}^{-1}}^2}_{\text{Term 3}}.
\label{eq:cost_function_regularised}
\end{equation}
{The cost function comprises three contributions.} Term~2 is the data-misfit term, which penalises the discrepancy between the model observables $\matr{M}_q\vect{\psi}$ and the measurements $\vect{d}$. The parameter regularisation is  twofold: 
Term~1 penalises excessive changes of parameters from the  previous iteration because  $\boldsymbol{\psi}^f$ includes $\boldsymbol{\alpha}^f$ (this is the standard prior that acts as an implicit regulariser on parameters when performing simultaneous state and parameter estimation);
and 
Term~3 is the contribution of the proposed PR-EnKF, which  discourages excessive changes of the parameters from their literature values, $\mathbf{p}$,  with uncertainty  $\mathbf{C}_{p\!p}$. 
The parameter pseudo-observation uncertainty, $\mathbf{C}_{p\!p}$, controls the strength of the parameter regularisation: a large $\matr{C}_{p\!p}$ (low confidence in $\vect{p}$) weakens {Term 3}. In the limit of infinite uncertainty, $\matr{C}_{p\!p}\rightarrow\infty$, the regularised cost function \eqref{eq:cost_function_regularised} tends to that of the EnKF \eqref{eq:cost_function_standard}.
}\\

The optimum $\vect{\psi}^a = \arg\min_{\vect{\psi}}\,\mathcal{J}(\vect{\psi})$ is the maximum \textit{a posteriori}  estimate that remains close to 
both the forecast and observations while {promoting updates close to} literature-consistent values $\vect{p}$. 
Setting $\mathrm{d}\mathcal{J}/\mathrm{d}\vect{\psi}_m = 0$ and grouping terms in $\vect{\psi}_m^a$ yields
\begin{equation}
\left(\matr{C}_{\psi\psi}^{f^{-1}} + \matr{M}_q^\top\matr{C}_{d\!d}^{-1}\matr{M}_q + \matr{M}_\alpha^\top\matr{C}_{p\!p}^{-1}\matr{M}_\alpha\right)\vect{\psi}_m^a = \matr{C}_{\psi\psi}^{f^{-1}}\vect{\psi}_m^f + \matr{M}_q^\top\matr{C}_{d\!d}^{-1}\vect{d}_m + \matr{M}_\alpha^\top\matr{C}_{p\!p}^{-1}\vect{p}_m.
\label{eq:grouped_system}
\end{equation}
To apply the Woodbury matrix identity \citep{Woodbury1950}, we define the block observation 
operator and covariance matrices as
\begin{equation}
\tilde{\matr{M}} = \begin{bmatrix} \matr{M}_q \\ \matr{M}_\alpha \end{bmatrix} \in \mathbb{R}^{N_\phi\times(N_q+N_\alpha)} \quad \text{and} \quad \tilde{\matr{C}} = \begin{bmatrix} \matr{C}_{d\!d} & \matr{0} \\ \matr{0} & \matr{C}_{p\!p} \end{bmatrix} \in \mathbb{R}^{(N_q+N_\alpha)\times(N_q+N_\alpha)},
\label{eq:combined_operators}
\end{equation}
such that the analysis update for each ensemble member $m = 1,\ldots,N_e$ becomes
\begin{equation}
\vect{\psi}_m^a = \vect{\psi}_m^f + \tilde{\matr{K}}\left(\begin{bmatrix}\vect{d}_m\\\vect{p}_m\end{bmatrix} - \tilde{\matr{M}}\vect{\psi}_m^f\right),
\label{eq:woodbury_update}
\end{equation}
where $\tilde{\matr{K}} = \matr{C}_{\psi\psi}^f\tilde{\matr{M}}^\top(\tilde{\matr{C}} + \tilde{\matr{M}}\matr{C}_{\psi\psi}^f\tilde{\matr{M}}^\top)^{-1}$ is the modified Kalman gain matrix accounting for both the observation and parameter regularisation terms. Analytically, the solution of the PR-EnKF yields the update for both states and parameters, respectively  
{
\begin{subequations}\label{eq:pr-EnKF}
\begin{equation}
    \begin{cases}
        \vect{\phi}_m^a = \vect{\phi}_m^f
            + {\matr{K}}_{\phi,d}
              (\vect{d}_m - \vect{q}_m^f)
            + {\matr{K}}_{\phi,p}
              (\vect{p}_m - \vect{\alpha}_m^f), \\[.8em]
        \vect{\alpha}_m^a = \vect{\alpha}_m^f
            + {\matr{K}}_{\alpha,d}
              (\vect{d}_m - \vect{q}_m^f)
            + {\matr{K}}_{\alpha,p}
              (\vect{p}_m - \vect{\alpha}_m^f),
    \end{cases}
    \label{eq:reg_enkf_state_param_update}
\end{equation}
where ${\matr{K}}_{\phi,d}$ and ${\matr{K}}_{\alpha,d}$ are identical to \eqref{eq:gains_enkf} and are present in ensemble methods for combined state and parameter estimation whenever $\matr{C}_{\alpha q}^f \neq \matr{0}$. 
 The additional matrices introduced with the proposed regularisation are
\begin{equation}
        {\matr{K}}_{\phi,p} = \matr{C}_{\phi\alpha}^f \left(\matr{C}_{p\!p}+\matr{C}_{\alpha\alpha}^f\right)^{-1}
        \quad \text{and} \quad
        {\matr{K}}_{\alpha,p} = \matr{C}_{\alpha\alpha}^f
        \left(\matr{C}_{p\!p}+\matr{C}_{\alpha\alpha}^f\right)^{-1}\!,
    \label{eq:gains_prenkf}
\end{equation}
\end{subequations}
which correct $\vect{\phi}$ and $\vect{\alpha}$ using  parameter innovations, according to  $\matr{C}_{\phi\alpha}^f$ and $\matr{C}_{\alpha\alpha}^f$, respectively.
The two gains \eqref{eq:gains_prenkf} vanish as $\matr{C}_{p\!p}^{-1}\to\matr{0}$; thus, \eqref{eq:reg_enkf_state_param_update} recovers the standard EnKF \eqref{eq:enkf_state_param_update}.
}
The prior knowledge on parameters embedded in the regularisation term affects the  state variables through the cross-covariance $\matr{C}_{\phi\alpha}^f$, yielding analysis states and parameters consistent with both observations $\vect{d}$ and literature-consistent parameter values $\vect{p}$. Unlike classical penalisation strategies \citep{Tikhonov1977}, where the penalty weight is fixed \textit{a priori}, {the regularisation contribution here is set by the gain $\matr{K}_{\alpha,p}$ \eqref{eq:gains_prenkf}, which weighs the fixed $\matr{C}_{pp}$ against the ensemble-estimated $\matr{C}_{\alpha\alpha}^f$. As the latter is updated each cycle (Algorithm~\ref{alg:prenkf_algorithm_detailed}), the regularisation strength adapts dynamically as the ensemble evolves, whilst preserving Bayesian consistency and optimality.}

The formulation above is written for a single observation field $\vect{q}$, but it extends directly to multiple observed quantities (e.g., velocity and turbulent kinetic energy) by augmenting the observation vector and the model state; this extension is described in \S\ref{sec:multifield}.

{
\subsection{Comment on computational efficiency}
\label{subsec:efficiency}
The combined operator $\tilde{\matr{M}}$ in \eqref{eq:combined_operators}
expands the measurement space from $N_q$ to $N_q + N_\alpha$. For typical urban
flow applications where $N_q \gg N_\alpha$, this expansion incurs a modest
increase in computational cost. {For the CEDVAL~A1-1 case considered here,
$N_q = 146$ and $N_\alpha = 11$, corresponding to a 24\% increase in inversion
cost.} Since the RANS solver dominates the overall computational budget and parameter updates occur only at discrete intervals, this per-step overhead represents a small fraction of the total cost. The improved convergence properties ( \S\ref{subsec:convergence_stability}) substantially reduce the total number of iterations required, resulting in a reduced overall computational time. The total computational cost of the PR-EnKF {inference} on the CEDVAL~A1-1 case amounts to approximately 9{,}000--10{,}000 CPU-hours on AMD EPYC 7542 cores.}

\section{Numerical methodology}
\label{sec:validation}
%

All simulations employ steady RANS equations with the SST $k$-$\omega$ turbulence closure model \citep{Menter1994SST}. {The  SST equations are included in Appendix \ref{app:sst} for completeness. }
The following subsections detail the atmospheric boundary layer (ABL) modelling and validation strategy.

\subsection{Atmospheric boundary layer modelling}
\label{Atmospheric}

Accurate atmospheric boundary layer representation is key to  urban flow simulations. We employ the ABL approach of \citet{Parente2011b} and \citet{Bellegoni},  which  guarantees streamwise homogeneity of inlet profiles throughout the domain, preventing spurious flow acceleration in undisturbed regions. 
The inlet streamwise velocity profile follows the logarithmic law
\begin{equation}
u(z) = \frac{u_*}{\kappa}\ln\left(\frac{z + z_0}{z_0}\right),
\label{eq:log_law}
\end{equation}
where $u_*$ is the friction velocity, $\kappa = 0.41$ is the von Kármán constant, and $z_0$ is the aerodynamic roughness length. 
For the inlet turbulent kinetic energy profile, $k$, we employ two formulations depending on experimental data characteristics. Monotonic profiles have a logarithmic law
\begin{equation}
k(z) = C_1\ln(z + z_0) + C_2,
\label{eq:k_log}
\end{equation}
where $C_1$ and $C_2$ are empirical constants. Complex urban terrains with a non-monotonic $k$ behaviour employ a four-parameter law 
\begin{equation}
k(z) = A\ln\left(\frac{z + z_0}{z_0}\right) + B\left(\frac{z + z_0}{z_0}\right)^2 + C\left(\frac{z + z_0}{z_0}\right) + D,
\label{eq:k_four_param}
\end{equation}
with coefficients $A$, $B$, $C$, and $D$ fitted to experimental profiles.  
In both cases, the specific dissipation rate follows the equilibrium assumption
$\omega(z) = {k(z)}/\left[{\kappa u_*(z + z_0)}\right]$.

Source terms in the $k$ and $\omega$ equations \citep{Parente2011b} and ABL-consistent wall functions \citep{Parente2011} maintain profile consistency in undisturbed regions. Near buildings, where equilibrium assumptions break down, we employ the following Building Influence Area (BIA) formulation, which smoothly transitions to the standard SST equations  \citep{Parente2011,Longo2017}
\begin{equation}
\text{BIA} = \max\left(\frac{|u - u_\text{ABL}|}{u_\text{ABL}}, \frac{|k - k_\text{ABL}|}{k_\text{ABL}}, \frac{|\omega - \omega_\text{ABL}|}{\omega_\text{ABL}}\right),
\label{eq:bia_definition}
\end{equation}
where the subscript ABL indicates values from the undisturbed inlet profiles. This hybrid approach enables the solution of the standard SST $k$-$\omega$ equations near buildings where complex flow phenomena require the full closure model, while maintaining the ABL formulation in the far field. 

\subsection{Test cases}

\label{subsec:test_setup}
\begin{figure}[h]
  \centerline{\includegraphics[width=0.9\textwidth]{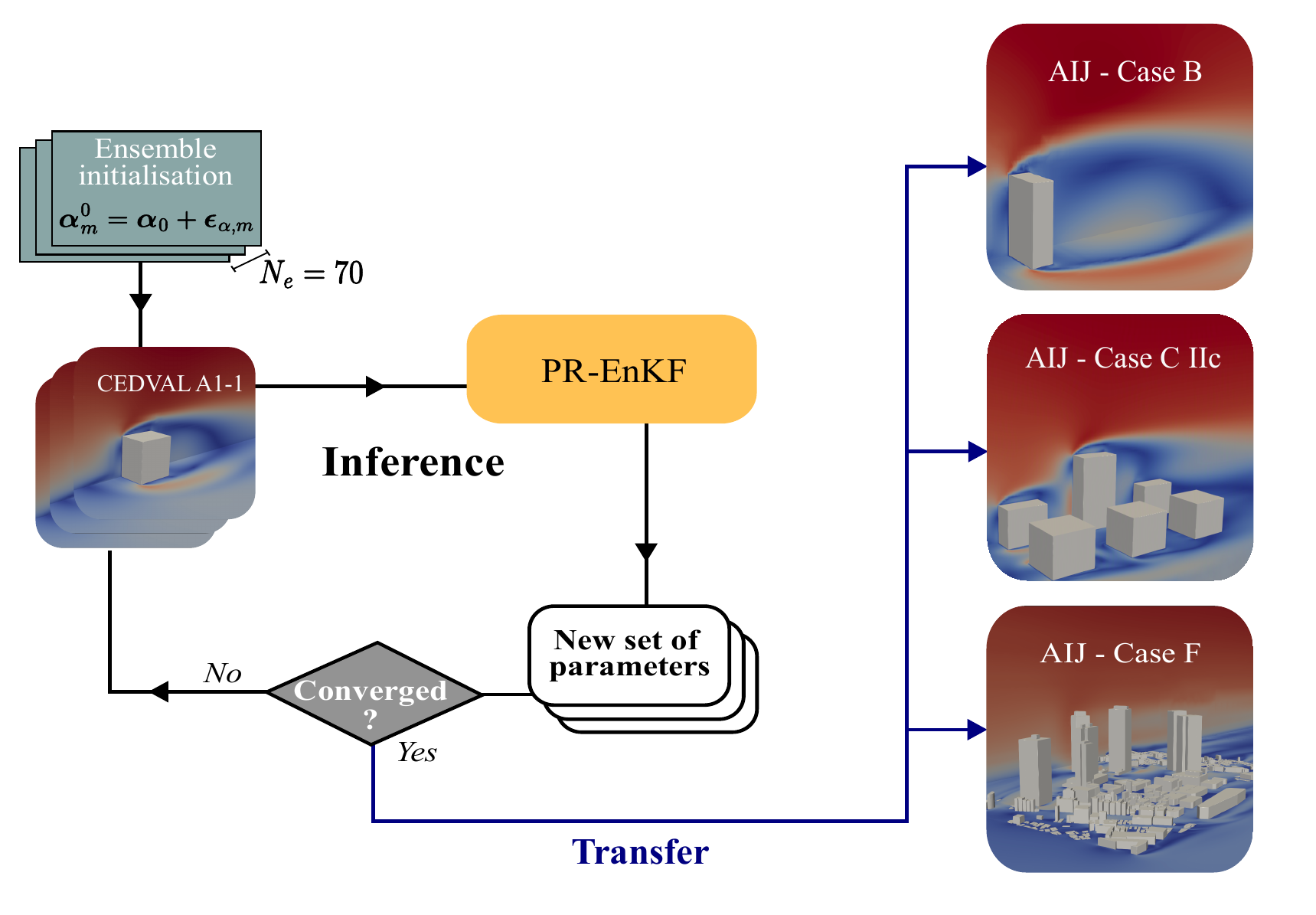}}
  \caption{{Schematic of the {assimilation}-transferability methodology. The PR-EnKF infers optimal SST $k-\omega$ parameters on the CEDVAL A1-1 isolated building. The optimal parameters are  subsequently transferred without modification to three cases of increasing complexity: the AIJ high-rise building (Case B), the building array (Case C), and the Shinjuku urban district (Case F).}}
  \label{fig:schematic}
\end{figure}
To assess the PR-EnKF's  transferability, we adopt an {assimilation}-transferability validation approach as described in Figure~\ref{fig:schematic}. First, the SST $k$-$\omega$ parameters are {inferred} on a computationally tractable case (CEDVAL A1-1 from \citet{CedvalHam}). Then, the {inferred} parameters are transferred without modification to increasingly complex geometries: a single high-rise building (Case B \citet{AIJ_CFD_Guide_2007}), an irregular building array (Case C \citet{AIJ_CFD_Guide_2007}), and the real-world Shinjuku urban district (Case F \citet{AIJ_CFD_Guide_2007}). 
All simulations use OpenFOAM's steady-state $p$-$U$ solver with second-order schemes, SIMPLE coupling, ABL wall functions \citep{Parente2011}, with computational meshes generated with  \texttt{snappyHexMesh}. {Simulation convergence is attained when the residuals stabilise below $10^{-6}$.}

\subsubsection{{Assimilation} case: CEDVAL A1-1 single building}\label{sec:opt_case}
The optimisation case features a single rectangular building (width 0.1~m, length 0.15~m, height $H=0.125$~m) from the CEDVAL database \citep{CedvalHam}. The computational domain extends 1~m upstream and 3~m downstream of the building, with the top boundary positioned sufficiently high to avoid blockage effects. 
Exploiting the symmetry plane at $y = 0$~m, only half of the domain is modelled. We employ a mesh with 1.7 million cells following a sensitivity analysis \citep{Bombardi2025}.

\begin{figure}[t]
  \centering
  \begin{tabular}{cc}
    \raisebox{-9pt}{\includegraphics[width=0.45\textwidth]{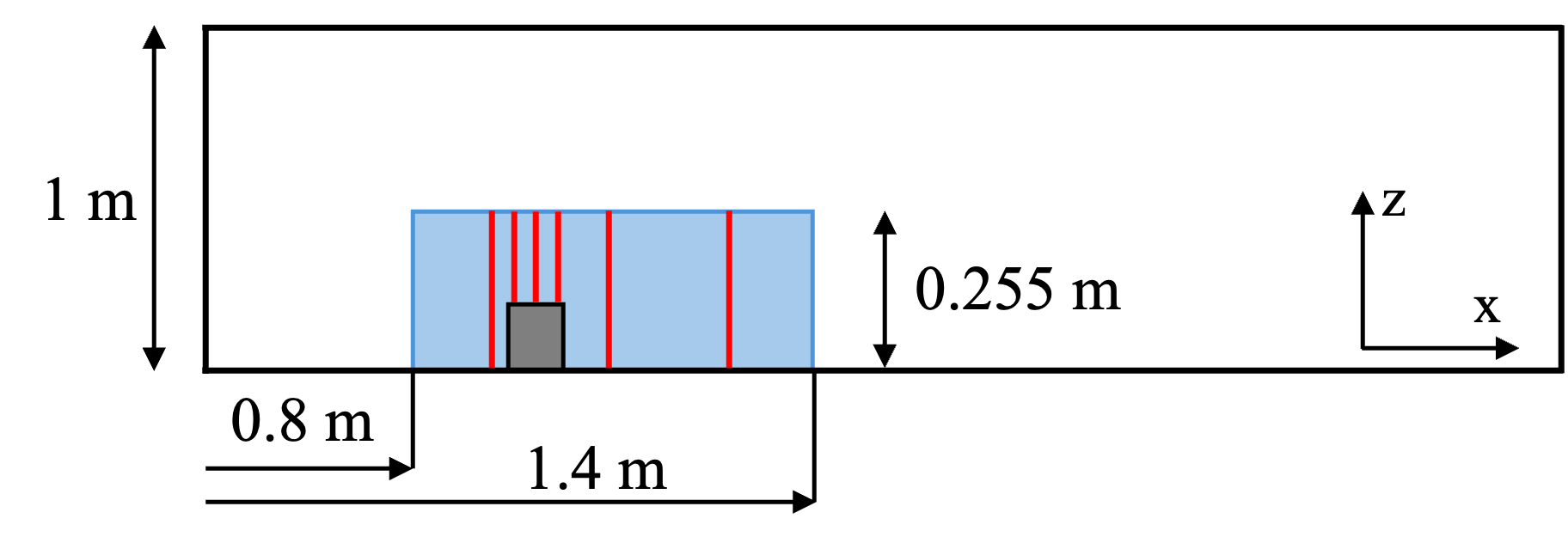}} &
    \raisebox{0pt}{\includegraphics[width=0.45\textwidth]{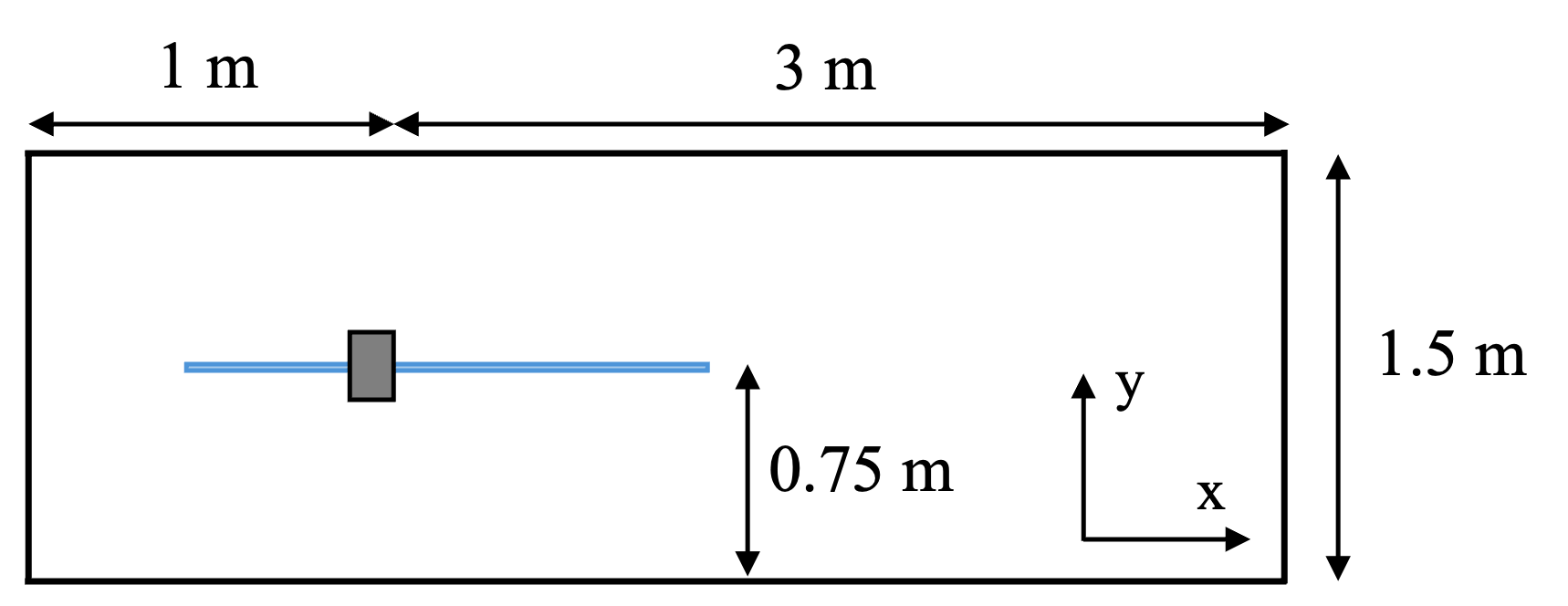}} \\
    (a) & (b)
  \end{tabular}
  \caption{Schematic representation of the CEDVAL A1-1 configuration showing \citep{CedvalHam} (\textit{a})  the mid-plane lateral view at $y=0$ with the computational domain dimensions and (\textit{b}) the corresponding top view. The blue-shaded region denotes the measurement area, with minimum and maximum coordinate limits indicated in metres. In panel (\textit{a}), the red lines mark the measurement locations at $x=\{-0.072, -0.04, 0, 0.04, 0.105, 0.3\}$ metres  from left to right.}
  \label{fig1}
\end{figure}

The experimental measurements used in data assimilation consist of velocity and turbulent kinetic energy data from wind tunnel tests at 6 locations distributed along the building, yielding $\approx$150 points (Figure~\ref{fig1}). 
The inlet profiles employ a logarithmic formulation for turbulent kinetic energy \eqref{eq:k_log}, with coefficients determined through  fitting to experimental  data. This case is the only configuration used to infer the parameters, and all other cases are used only for the transferability assessment. 
%
%

\subsubsection{Transferability cases}\label{sec:transfer_cases}

\begin{figure}[t]
  \centering
  \begin{tabular}{cc}
    \raisebox{-8pt}{\includegraphics[width=0.45\textwidth]{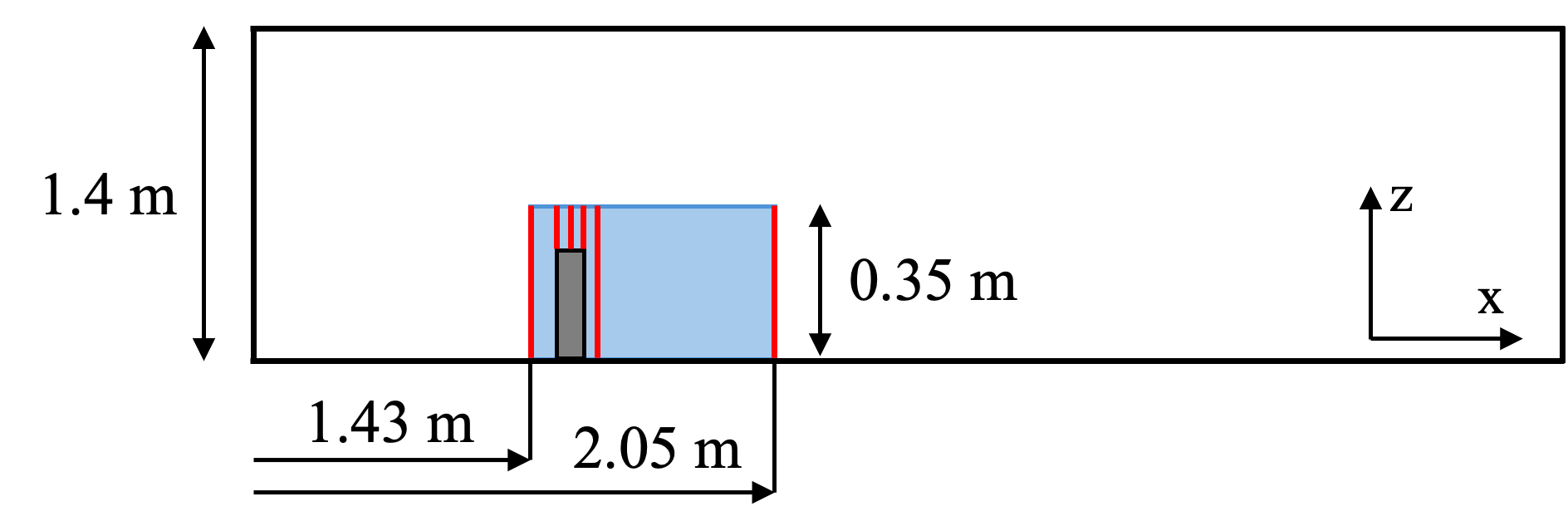}} &
    \raisebox{0pt}{\includegraphics[width=0.45\textwidth]{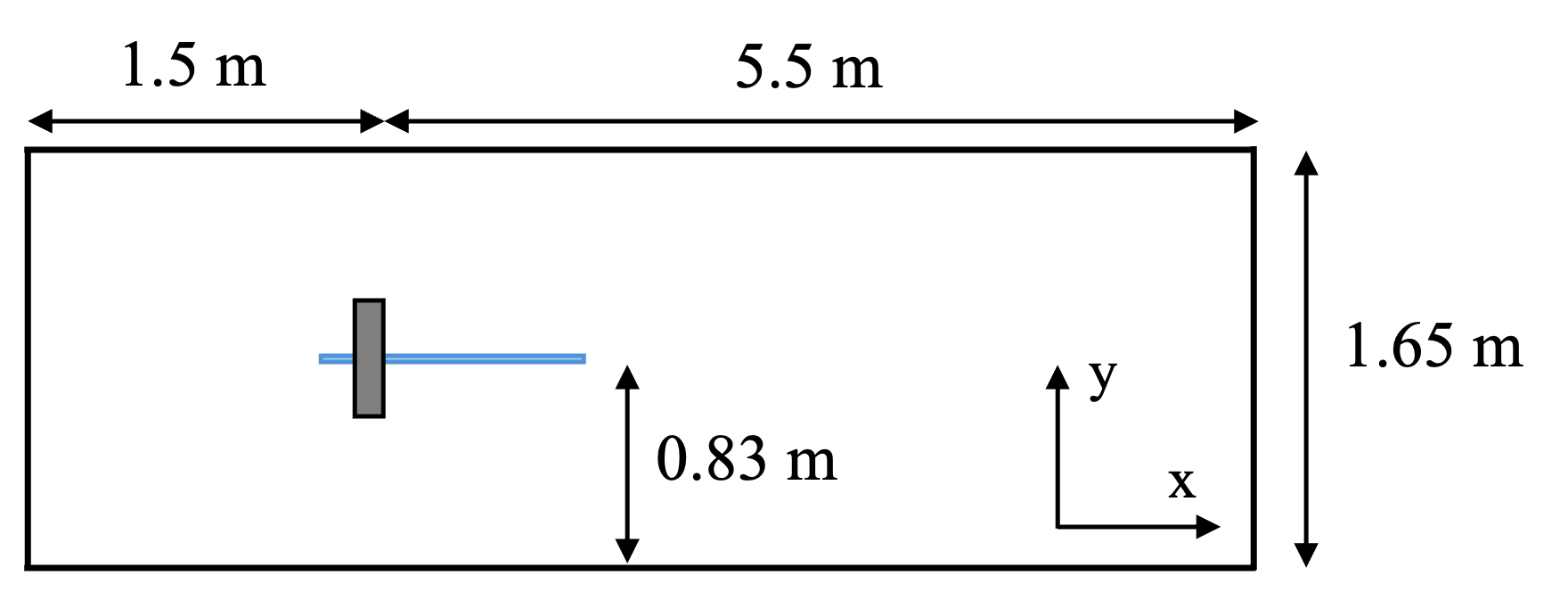}} \\
    \multicolumn{2}{c}{(a)} \\[1ex]
    \raisebox{3.5pt}{\includegraphics[width=0.45\textwidth]{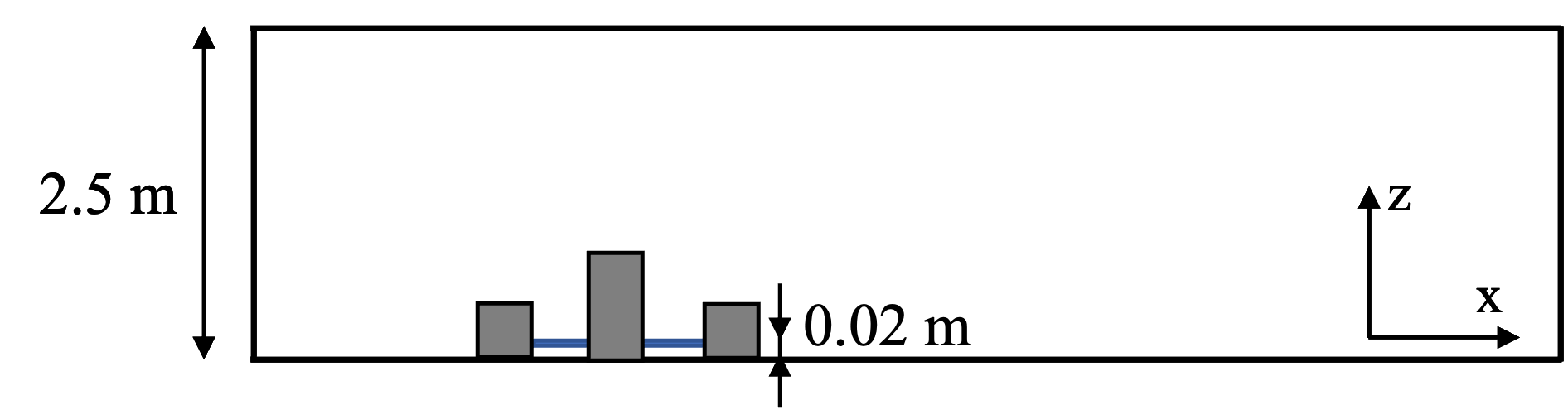}} &
    \raisebox{0pt}{\includegraphics[width=0.45\textwidth]{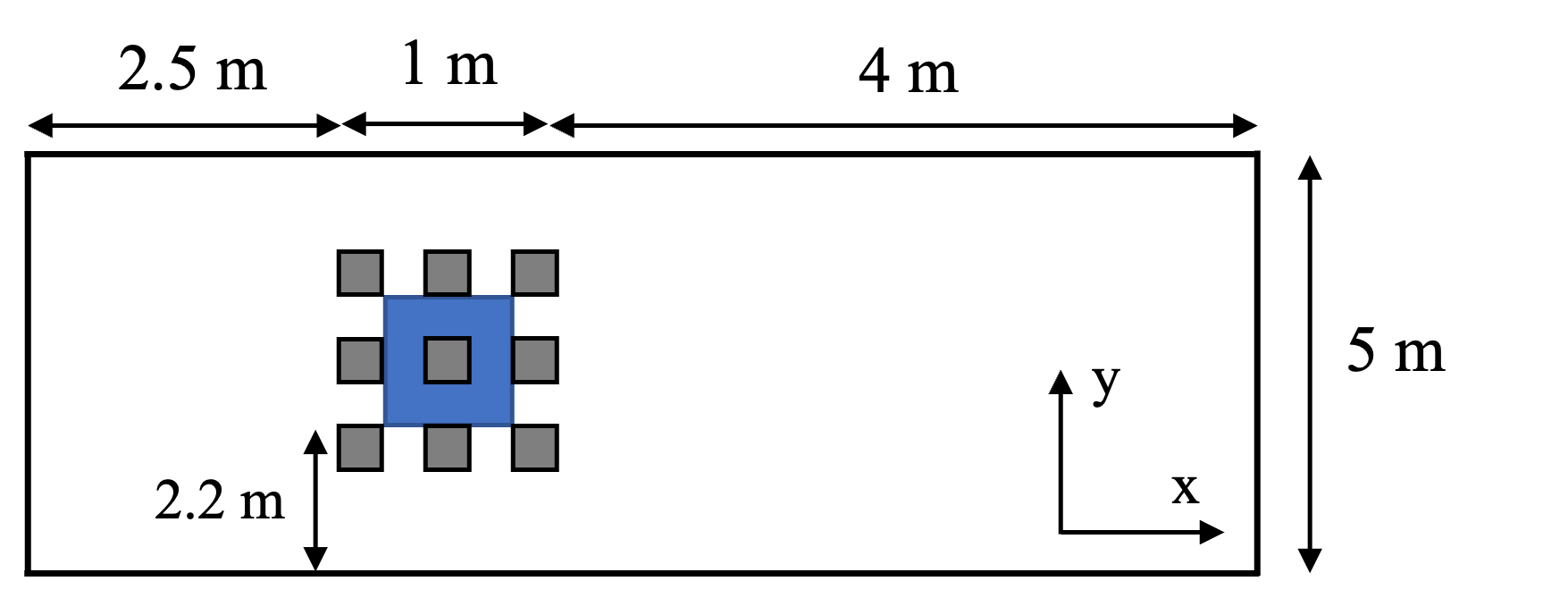}} \\
    \multicolumn{2}{c}{(b)} \\[1ex]
    \multicolumn{2}{c}{\raisebox{0pt}{\includegraphics[width=0.8\textwidth]{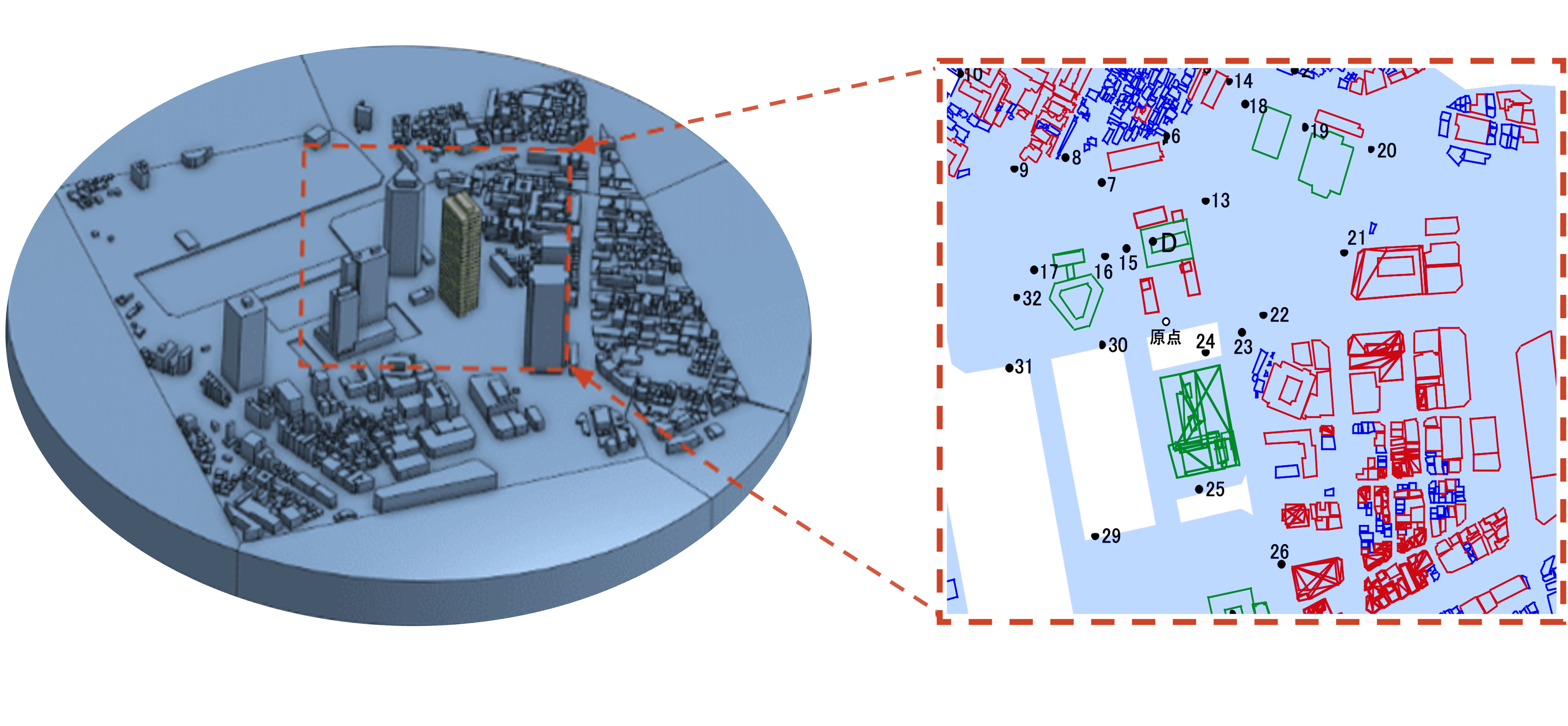}}} \\
    \multicolumn{2}{c}{(c)}
  \end{tabular}
  \caption{
  Schematic representation of the three transferability cases from \citep{AIJ_CFD_Guide_2007}. 
  (a)   AIJ Case B : mid-plane lateral view at $y=0$ and top view, where the measurement region is  shaded in blue and the red lines mark the measurement planes at  $x=\{-0.075, -0.025, 0, 0.025, 0.05, 0.55\}$ metres. 
  (b) AIJ Case C IIc: mid-plane lateral view at $y=0$  and top view, where the measurement area at  $z = 0.02$ m is shaded in blue. 
   (c)  Shinjuku district Case F:  3D view of the urban geometry within cylindrical computational domain, and plan view showing the measurement locations at $z = 10$ (numbered points), with buildings colour-coded by height. 
  }\label{fig:transfer_cases}
\end{figure}
Figure~\ref{fig:transfer_cases} summarises the three transferability cases used to evaluate whether the PR-EnKF parameters from one  {assimilation} setting can generalise to increasingly complex urban flows. The cases span a single isolated building, a building array, and a full urban district, which are explained below. \\ 

\begin{description}
\item[\textit{High-rise building  \citep[Case B in][]{AIJ_CFD_Guide_2007}}.] A single high-rise building with height and width four times its depth. The measurement data comprise velocity and kinetic energy vertical profiles at six locations  (Figure~\ref{fig:transfer_cases}a). 
Similarly to \S\ref{sec:opt_case},  only half the domain is modelled because of the symmetry in $y$. The mesh contains approximately 1.8 million cells, the inlet conditions use $z_0=9.6\times10^{-5}$~m, $u_*=0.302$~m/s \citep{Yoshihide2004}, and the four-parameter $k$ profile \eqref{eq:k_four_param} has  $A = 0.065$, $B = 1.25 \times 10^{-9}$, $C = -5.9 \times 10^{-5}$, and $D = 0.23$. 
The flow physics remain  qualitatively similar to the optimisation case but the larger aspect ratio produces stronger three-dimensional wake and downwash effects. 

\item[\textit{Building array  \citep[Case C][]{AIJ_CFD_Guide_2007}}.] A $3\times3$ array with a central building of height $2H$ and surrounding buildings of height $H=0.2$~m. We assimilate velocity measurements from  $\approx80$ datapoints distributed in a plane at $z=0.02$~m under normal incidence   (Figure~\ref{fig:transfer_cases}b). The inlet boundary condition for turbulent kinetic energy employs the increasing-decreasing four-parameter formulation  \eqref{eq:k_four_param} and the  mesh contains approximately 3.5 million cells. This case introduces wake interaction and channelling between buildings.

\item[\textit{Urban district \citep[Shinjuku Case F][]{AIJ_CFD_Guide_2007}.}] This realistic test case examines the flow field within the Shinjuku sub-central area of Tokyo, Japan.
The domain spans $1~\text{km}\times1~\text{km}$ and contains high-rise buildings with highly irregular geometry, including the Shinjuku Mitsui Building (225~m). A cylindrical computational domain of radius 3.5~km and height 1350~m is discretised with approximately 32 million hexahedral cells. 
The mesh resolution is finest near walls and the ground, with horizontal and lateral resolution of 1.5~m and vertical resolution of 0.5~m. Above this region, the maximum cell size increases to 18~m in all three directions, and further increases to 36~m above 300~m. 
The measurement data contain velocity data at 31 locations throughout the district at 10~m height, with two exceptions: point C at the KDD building top (192~m) and point D at the SMB top (242~m). The simulated velocities are normalised by the predicted velocity at point C. Some of these points are numbered in the close-up in Figure~\ref{fig:transfer_cases}c.  %
The inlet roughness length is $z_0=0.85$~m, and the four-parameter turbulent kinetic energy profile is fitted with $A = 0.374$, $B = 5.86 \times 10^{-6}$, $C = -0.0079$, and $D = 0.48$. This full-scale urban environment is the most challenging test for  parameter transferability. 

\end{description}

\subsection{Implementation of the PR-EnKF}
\label{sec:implementation}

The complete PR-EnKF workflow is summarised in Algorithm~\ref{alg:prenkf_algorithm_detailed}. 
The following subsections detail the selection of the literature-consistent parameters (\S\ref{subsec:prior_specification}), the multi-field assimilation strategy (\S\ref{sec:multifield}), and the ensemble initialisation and inflation used in data assimilation (\S\ref{sec:ensemble_init_inflate}).

\subsubsection{Parameter selection and regularisation}
\label{subsec:prior_specification}

We {infer} the 11 parameters of the Shear Stress Transport (SST) $k$-$\omega$ turbulence model \citep{Menter1994SST}. The SST model is a two-equation RANS closure that blends the standard $k$-$\epsilon$ and $k$-$\omega$ formulations \citep{LaunderSpalding1974, Wilcox1988} and is well suited to separated flows and adverse pressure gradients. {Table~\ref{tab:sst_parameters} lists these parameters and their original values from \citet{Menter1994SST}, which are denoted as the default values and used in the unmodified baseline CFD simulations. In the PR-EnKF, these default values correspond to the literature-consistent parameters $\vect{p}$ (Term 3 in \eqref{eq:cost_function_regularised}).}
\begin{table}
  \begin{center}
  \def~{\hphantom{0}}
  \begin{tabular}{lll}
    \hline
    Parameter & Default Value & Description \\
    \hline
    $a_1$ & 0.31 & Eddy viscosity coefficient \\
    $b_1$ & 1.0 & Blending coefficient \\
    $c_1$ & 10.0 & Production limiter \\
    $\beta^*$ & 0.09 & Turbulence destruction coefficient \\
    $\alpha_{k_1}$ & 0.85 & $k$-$\omega$ diffusion coefficient for $k$ \\
    $\alpha_{k_2}$ & 1.0 & $k$-$\epsilon$ diffusion coefficient for $k$ \\
    $\alpha_{\omega_1}$ & 0.5 & $k$-$\omega$ diffusion coefficient for $\omega$ \\
    $\alpha_{\omega_2}$ & 0.856 & $k$-$\epsilon$ diffusion coefficient for $\omega$ \\
    $\gamma_1$ & 0.553 & $k$-$\omega$ production coefficient \\
    $\gamma_2$ & 0.44 & $k$-$\epsilon$ production coefficient \\
    $\beta_2$ & 0.0828 & $k$-$\epsilon$ destruction coefficient \\
    \hline
  \end{tabular}
  \caption{{SST $k$-$\omega$ turbulence model parameters and their default values from \citet{Menter1994SST}, used as the literature-consistent values $\vect{p}$ for regularisation (Term 3 in \eqref{eq:cost_function_regularised}).}}
  \label{tab:sst_parameters}
  \end{center}
\end{table}
%
%
The covariance matrix of the literature-consistent values $\vect{p}$ is pre-defined as
\begin{equation}
{C}_{p\!p,ii} = \left(\sigma_i \cdot p_i\right)^2,
\label{eq:prior_cov}
\end{equation}
where $p_i$ is the literature-consistent value of the $i$th parameter and $\sigma_i$ is a user-defined uncertainty. This choice scales the prior uncertainty with the parameter magnitude and keeps the regularisation dimensionally consistent.  
For simplicity, we use a uniform uncertainty for all parameters.
We find that a 20\% uncertainty provides enough flexibility to adjust {the parameters that need correction, while leaving those already close to the optimum near their literature values (\S\ref{subsec:convergence_stability}).}

{
\subsubsection{Multi-field data assimilation}
\label{sec:multifield}

The simultaneous assimilation of observations from multiple physical quantities is referred to as ``multi-field {assimilation}''. Assimilating heterogeneous observations can improve combined state and parameter estimation, as shown by \citet{ZhangCF2021}. In this work, the additional observations are the velocity and turbulent kinetic energy (TKE) for each ensemble member (Algorithm \eqref{alg:prenkf_algorithm_detailed}) 
\begin{equation}
\vect{d}_m^* = [\vect{d}_{U,m}; \vect{d}_{k,m}],
\end{equation}
where subscripts $U$ and $k$ denote velocity and TKE, respectively. 
To assimilate both sources of information, we augment the state vector and measurement operators as
\begin{equation}
\vect{\psi}^* =
\begin{bmatrix}
\vect{\phi}_U, \vect{\phi}_k, \vect{\alpha}, \vect{q}_U, \vect{q}_k
\end{bmatrix}^\top
\in \mathbb{R}^{N_{\phi_U} + N_{\phi_k} + N_\alpha + N_{q_U} + N_{q_k}},
\label{eq:multifield_state}
\end{equation}
where $\vect{\phi}_U$ and $\vect{\phi}_k$ are the state variables associated with the velocity and TKE fields, and $\vect{q}_U$ and $\vect{q}_k$ are the corresponding model observables. The combined observation operator and covariance become
\begin{equation}
\tilde{\matr{M}}^* =
\begin{bmatrix}
\matr{M}_U \\
\matr{M}_k \\
\matr{M}_\alpha
\end{bmatrix},
\qquad
\tilde{\matr{C}}^* =
\begin{bmatrix}
\matr{C}_{dd,U} & \matr{0} & \matr{0} \\
\matr{0}        & \matr{C}_{dd,k} & \matr{0} \\
\matr{0}        & \matr{0}        & \matr{C}_{p\!p}
\end{bmatrix},
\label{eq:multifield_combined}
\end{equation}
where $\matr{M}_U$ and $\matr{M}_k$ map the state vector to velocity and TKE observations, and $\matr{C}_{dd,U}$ and $\matr{C}_{dd,k}$ are the corresponding observation error covariances. The PR-EnKF update in \S\ref{subsec:regularised_enkf} is then applied with an augmented covariance matrix that includes the cross-covariances between the parameters and each data field, and between the fields themselves. Parameters that correlate strongly with TKE but weakly with velocity are primarily informed by TKE observations, and vice versa, so each observation field mainly updates the parameters to which it is most sensitive.

Lastly, because velocity and TKE have different numerical ranges, practical implementation requires normalisation to avoid one field dominating the update. We apply a min-max normalisation mapping each field to $[0,1]$ before computing covariances and performing the analysis step, and rescale all quantities back to physical units after the PR-EnKF update.

\subsubsection{Ensemble initialisation and inflation}
\label{sec:ensemble_init_inflate}

The ensemble  is initialised by perturbing the model parameters around their literature-consistent values as
\begin{equation}
\vect{\alpha}_{m}^0 = \vect{\alpha}_0 + \vect{\epsilon}_{\alpha,m}, \quad \text{for} \quad m=1\dots,N_e
\end{equation}
with $\vect{\epsilon}_{\alpha,m} \sim \mathcal{N}(\vect{0}, \matr{C}_{\alpha\alpha}^0)$, where we set $\matr{C}_{\alpha\alpha}^0 = \matr{C}_{p\!p}$ as defined in~\eqref{eq:prior_cov}.  
The ensemble size is set to $N_e = 70$. Parameters change by less than 5\% between $N_e=70$ and $N_e=90$. Because finite ensembles can lead to covariance underestimation and collapse \citep{Anderson2001}, we apply multiplicative covariance inflation after each analysis step as
\begin{equation}
\vect{\psi}_m^a \leftarrow \bar{\vect{\psi}}^a + \lambda(\vect{\psi}_m^a - \bar{\vect{\psi}}^a),
\end{equation}
with inflation factor $\lambda = 1.1$, which prevents covariance collapse whilst maintaining accuracy. (We performed numerical tests with $\lambda \in \{1.05, 1.1, 1.2\}$, and found small differences in the performance---result not shown.) 
}

\begin{algorithm}
\SetAlgoLined

\KwIn{Observations $\vect{d}$, literature-consistent parameters $\vect{p}$, RANS initial conditions 
$\vect{\phi}_0, \vect{\alpha}_0$, and error matrices $\matr{C}_{\alpha\alpha}^0$, $\matr{C}_{d\!d}$, and $\matr{C}_{p\!p}$}
\KwOut{{Inferred} parameters $\{\vect{\alpha}_m^a\}_{m=1}^{N_e}$ and states $\{\vect{\phi}_m^a\}_{m=1}^{N_e}$}

\BlankLine
\tcp{\small Initialise ensemble with $\vect{\epsilon}_{\alpha,m}\sim \mathcal{N}(\vect{0},\matr{C}_{\alpha\alpha}^0)$  }
$\{\vect{\alpha}_m\}_{m=1}^{N_e} \leftarrow \vect{\alpha}_0\mathbf{1}^\top + \{\vect{\epsilon}_{\alpha,m}\}_{m=1}^{N_e}$\\
$\{\vect{\phi}_m\}_{m=1}^{N_e} \leftarrow [\vect{\phi}_{1,0},\ldots,\vect{\phi}_{N_e,0}]$ 
\BlankLine

\tcp{\small Perturb (pseudo)observations with $\vect{\epsilon}_{p,m}\sim \mathcal{N}(\vect{0},\matr{C}_{p\!p}), \vect{\epsilon}_{d,m}\sim \mathcal{N}(\vect{0},\matr{C}_{d\!d})$}
$\matr{P} \leftarrow \{\vect{p} + \vect{\epsilon}_{p,m}\}_{m=1}^{N_e}$\\
$\matr{D} \leftarrow \{\vect{d} + \vect{\epsilon}_{d,m}\}_{m=1}^{N_e}$ \textit{\small //$\quad$MF: concatenate $\vect{d}^* = [\vect{d}_{U};\vect{d}_{k}]$} 

\BlankLine

\While{not converged}{
    \tcp{ Advance ensemble of RANS solvers for $\Delta n_a$ steps}
    \For{$m = 1,\ldots,N_e$}{
        $\vect{\phi}_{m}^f \leftarrow \mathcal{F}^{(\Delta n_a)}(\vect{\phi}_{m}, \vect{\alpha}_{m})$\\
        $\vect{\alpha}_{m}^f \leftarrow \vect{\alpha}_{m}$\\
        $\vect{q}_{m}^f \leftarrow \mathcal{M}(\vect{x}, \vect{\phi}_{m}^f)$ \textit{\small //$\quad$MF: form $\vect{q}_{m}^f = [\vect{q}_{U,m}^f;\vect{q}_{k,m}^f]$ as in \S\ref{sec:multifield}}
    }

     \tcp{ Build augmented state and estimate covariances }
    $\{\vect{\psi}_{m}^f\}_{m=1}^{N_e} \leftarrow
    \left\{\![\vect{\phi}_{m}^f;\vect{\alpha}_{m}^f;\vect{q}_{m}^f]^\top\!\right\}_{m=1}^{N_e}$ \\
    \textit{MF}: Apply min-max normalisation.\\

    $\mathcal{C} \leftarrow \big[
        \matr{C}_{\phi q}^f,\;
        \matr{C}_{\alpha q}^f,\;
        \matr{C}_{qq}^f,\;
        \matr{C}_{\phi\alpha}^f,\;
        \matr{C}_{\alpha\alpha}^f\big]$ \eqref{eq:ensemble_cov}  \textit{\small //$\;$MF: include $\matr{C}_{\alpha q_U}^f$, $\matr{C}_{\alpha q_k}^f$, $\matr{C}_{q_U q_k}^f$}

\BlankLine
	\tcp{ Analysis step and inflation}
    $\{\vect{\psi}_{m}^a\}_{m=1}^{N_e} \leftarrow
    \text{PR-EnKF}\bigl(
        \{\vect{\psi}_{m}^f\}_{m=1}^{N_e},
        \matr{D},
        \matr{P},
        \mathcal{C}
    \bigr)$\;

    $\{\vect{\psi}_{m}^a\}_{m=1}^{N_e} \leftarrow
    \bar{\vect{\psi}}^a +
    \lambda\left(\{\vect{\psi}_{m}^a\}_{m=1}^{N_e} - \bar{\vect{\psi}}^a\right)$\;
    \textit{MF}: Denormalise $\{\vect{\psi}_{m}^a\}_{m=1}^{N_e}$ back to physical units.
\BlankLine

\BlankLine
	\tcp{ Reinitialise RANS solvers}
    \For{$m = 1,\ldots,N_e$}{

    $\vect{\phi}_m  \leftarrow
    \matr{M}_\phi\vect{\psi}_{m}^a$ \\$  \vect{\alpha}_m \leftarrow \matr{M}_\alpha\vect{\psi}_{m}^a$\;
}
}

\caption{Parameter-Regularised Ensemble Kalman Filter pseudo-algorithm. \textit{MF} indicates additional steps for the multi-field assimilation.}
\label{alg:prenkf_algorithm_detailed}
\end{algorithm}

\section{Results: {Inference} of RANS parameters via the PR-EnKF}
\label{sec:results}

This section presents the PR-EnKF results for the CEDVAL A1-1 case \citep{CedvalHam}, which is the only case used for parameter inference. Parameter updates are applied at intervals of $\Delta n_a = 500$ RANS iterations. {(Shorter intervals update the parameters before the flow has responded, yielding noisier cross-covariance estimates, whereas longer intervals have increased computational cost and less effect on the update.)} The {inference} is run for 40 EnKF cycles, corresponding to 20,000 steady RANS iterations per ensemble member. For $N_e = 70$, {this amounts to a total of $1.4\times10^{6}$ RANS iterations across the ensemble,} equivalent to about 200 fully converged RANS simulations in  the same domain. No early termination criterion is imposed; therefore, long-term stability and solution drifts can be assessed. \\

In the upcoming subsections, we compare the performance of the PR-EnKF to the standard EnKF (\S\ref{subsec:standard_enkf}). 
First, \S\ref{subsec:convergence_stability} examines the convergence and stability of the two approaches using velocity-only observations, and the contribution of the regularisation to the analysis parameters.  
Second, \S\ref{subsec:tke_extension} extends the analysis to the multi-field assimilation of velocity and turbulent kinetic energy.
Third, we investigate the accuracy of the predicted velocity and TKE fields using the optimised parameters in \S\ref{subsec:accuracy}. 
 Table~\ref{tab:optimized_parameters} shows the mean and standard deviation of the 11 SST parameters obtained after  40 assimilation cycles for all the assimilation cases included in this section.

%

\begin{table}
\begin{center}
\begin{small}

\begin{tabular}{lccccc}
 &  &  &  \multicolumn{3}{c}{PR-EnKF } \\ \cline{4-6} 
\multicolumn{1}{c}{Parameter} & Default {value}&EnKF  &$\begin{array}{c} \sigma=0.1 \\ u\text{-only (\S\ref{subsec:convergence_stability})} \end{array}$&  $\begin{array}{c} \sigma=0.2 \\ u\text{-only (\S\ref{subsec:convergence_stability})} \end{array}$  & 
$\begin{array}{c} \sigma=0.2 \\ u\&k\text{ (\S\ref{subsec:tke_extension})} \end{array}$\\ \hline
$a_1$ & 0.31 & $1.480 \pm 0.366$ & $0.298 \pm 0.015$ & $0.294 \pm 0.026$ & $0.353 \pm 0.031$ \\
\rowcolor{cyan!9}$\alpha_{\omega2}$ & 0.856 & $0.593 \pm 0.409$ & $0.862 \pm 0.048$ & $0.865 \pm 0.099$ & $0.440 \pm 0.019$ \\
\rowcolor{cyan!9}$\beta^*$ & 0.09 & $0.087 \pm 0.041$ & $0.062 \pm 0.005$ & $0.066 \pm 0.009$ & $0.158 \pm 0.016$ \\
$b_1$ & 1.0 & $5.038 \pm 0.459$ & $1.020 \pm 0.053$ & $1.035 \pm 0.092$ & $1.130 \pm 0.097$ \\
$c_1$ & 10 & $7.469 \pm 3.651$ & $10.031 \pm 0.597$ & $10.066 \pm 1.170$ & $10.256 \pm 1.223$ \\
\rowcolor{cyan!9}$\alpha_{k1}$ & 0.85 & $0.303 \pm 0.253$ & $0.841 \pm 0.045$ & $0.794 \pm 0.103$ & $0.382 \pm 0.090$ \\
$\alpha_{k2}$ & 1.0 & $0.051 \pm 0.144$ & $1.010 \pm 0.058$ & $0.905 \pm 0.115$ & $0.956 \pm 0.203$ \\
\rowcolor{cyan!9}$\alpha_{\omega1}$ & 0.50 & $1.654 \pm 0.489$ & $0.501 \pm 0.027$ & $0.517 \pm 0.060$ & $0.298 \pm 0.027$ \\
\rowcolor{cyan!9}$\gamma_1$ & 0.555 & $0.136 \pm 0.085$ & $0.537 \pm 0.033$ & $0.471 \pm 0.061$ & $0.193 \pm 0.029$ \\
\rowcolor{cyan!9}$\gamma_2$ & 0.44 & $0.096 \pm 0.078$ & $0.437 \pm 0.023$ & $0.406 \pm 0.055$ & $0.132 \pm 0.049$ \\
\rowcolor{cyan!9}$\beta_2$ & 0.0828 & $0.128 \pm 0.084$ & $0.090 \pm 0.006$ & $0.086 \pm 0.010$ & $0.127 \pm 0.015$ \\ \hline
$u$-RMSE (m/s) & 0.998 & 0.485 & 0.521 & 0.504 & 0.516 \\
$k$-RMSE (m$^2$/s$^2$) & 1.421 & 1.023 & 0.961 & 0.938 & 0.881 \\ \hline
\end{tabular}
  \caption{Comparison of the optimised $k$-$\omega$ SST turbulence model parameters with the EnKF and different PR-EnKF configurations. 
  {Default values are the original SST coefficients of \citet{Menter1994SST} (Table~\ref{tab:sst_parameters}), used here as the literature-consistent values $\vect{p}$}. The highlighted rows (blue) indicate parameters showing sensitivity to the turbulent kinetic energy measurements. Bottom rows show the velocity and TKE RMSE on the CEDVAL A1-1 case.
  }
  \label{tab:optimized_parameters}
  \end{small}
  \end{center}
\end{table}

\subsection{Velocity-only assimilation}
\label{subsec:convergence_stability}

\begin{figure}[h]
  \centerline{\includegraphics[width=0.99\textwidth]{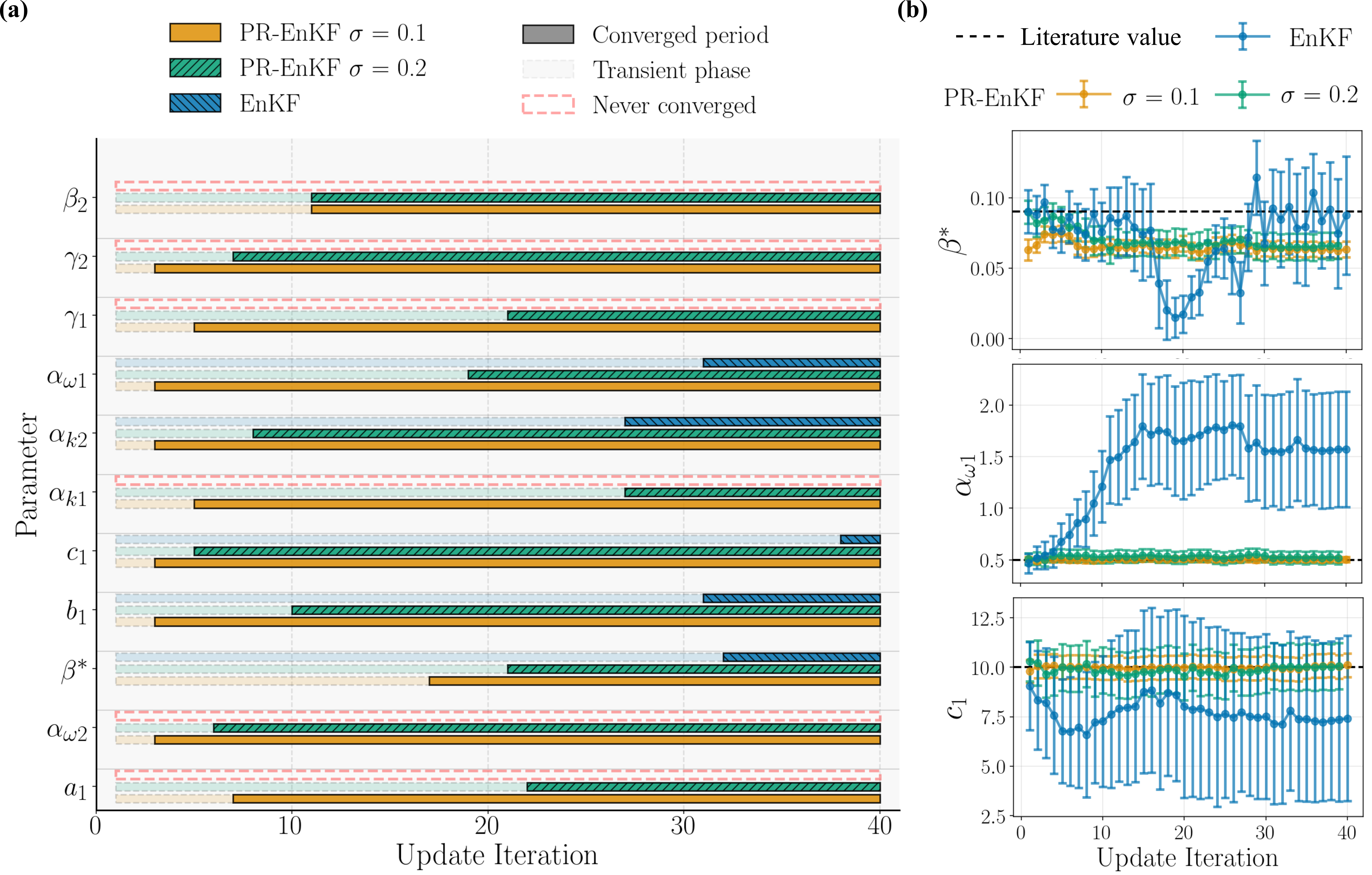}}
\caption{Parameter convergence with the standard EnKF, PR-EnKF with $\sigma=0.1$, and PR-EnKF with $\sigma=0.2$. (a) Convergence for all 11 SST $k$-$\omega$ parameters. Solid bars indicate the converged period, hatched bars the transient phase, and dashed boxes parameters that did not converge within the 40-iteration window. (b) Detailed convergence of three representative parameters ($\beta^*$, $\alpha_{\omega1}$, and $c_1$), showing the ensemble mean and standard deviation over 40 update iterations. The dashed lines indicate the literature-consistent values.}
  \label{fig:convergence}
\end{figure}

Figure~\ref{fig:convergence} compares the convergence behaviour of the standard EnKF against the PR-EnKF variants ($\sigma=0.1$ and $\sigma=0.2$). Specifically, Figure~\ref{fig:convergence}a displays the convergence  for the 11 SST parameters, where a parameter is considered converged when the relative difference between its current value and the mean of the last five iterations falls below 2\%. The PR-EnKF successfully converges for all 11 parameters within 20--30 iterations, with the stronger regularisation case (i.e., $\sigma = 0.1$) converging the fastest. In contrast, the standard EnKF is not stable, with a few parameters failing to converge in a 40-iteration window. 
%
The ensemble history in Figure~\ref{fig:convergence}b  shows the lack of convergence for the standard EnKF, which yields ensemble spreads on the order of  50--100\% for $\beta^*$, $\alpha_{\omega1}$, and $c_1$, whereas both regularised variants remain bounded and converge rapidly.
The regularised configurations produce parameters within the literature-consistent ranges: PR-EnKF with $\sigma=0.1$ averages 5.93\% ensemble spread (uncertainty), while the PR-EnKF with $\sigma=0.2$ averages 11.90\%. 
The standard EnKF produces  different results: the average ensemble spread is 101.58\%, and the mean values deviate from their initial guesses up to  377\% for  $a_1$ or 404\%  for $b_1$. 
Beyond 40 assimilation cycles, the  EnKF does not converge and its ensemble spread continues to grow (result not shown).  Hereafter, we employ $\sigma=0.2$ unless stated otherwise. 
\\

\begin{figure}[h]
  \centerline{\includegraphics[width=0.95\textwidth]{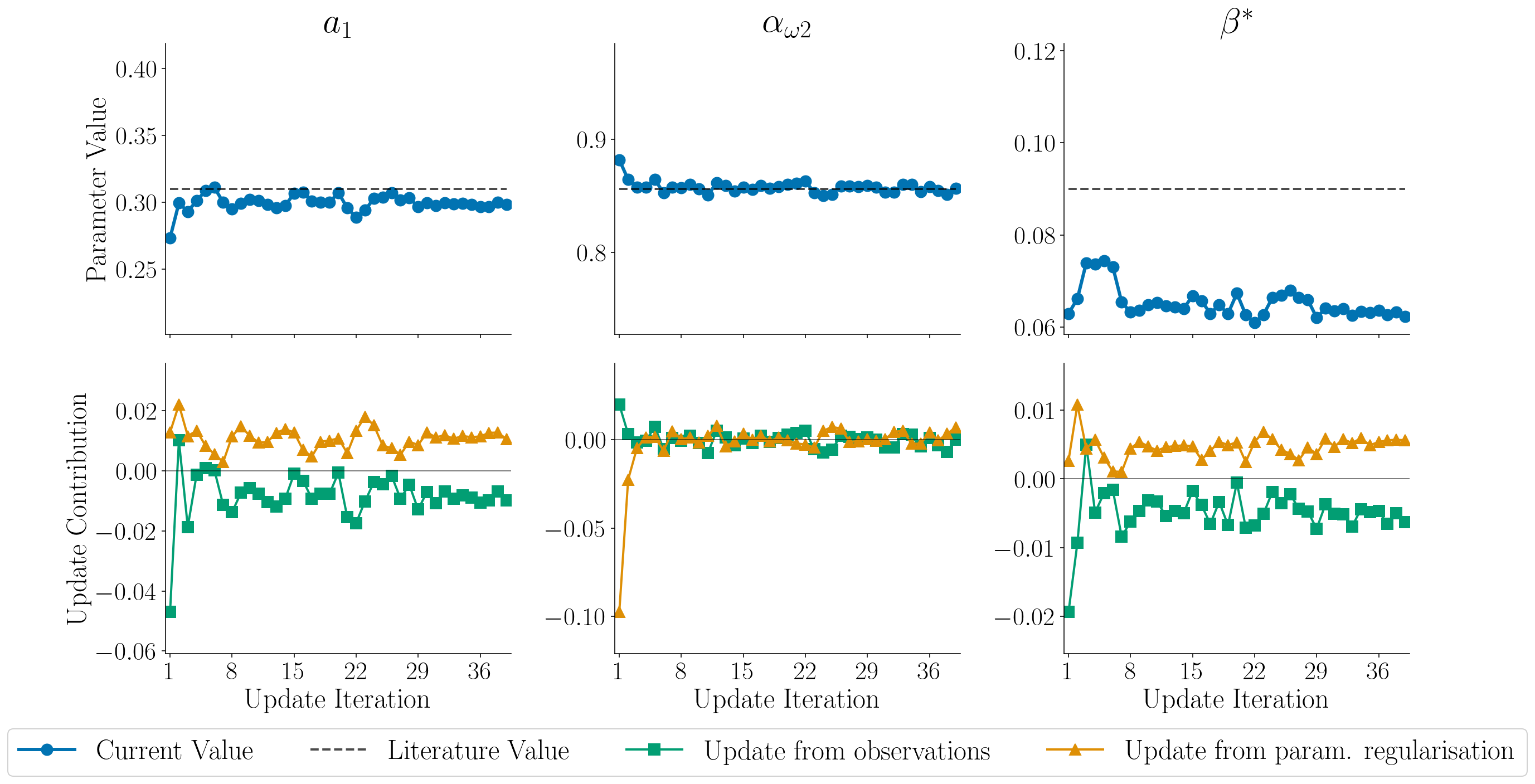}}
  \caption{
  Contribution from the observations and parameter regularisation to the mean analysis for three turbulence model parameters ($a_1$, $\alpha_{\omega_2}$, $\beta^*$). The top row shows the ensemble mean convergence and the literature-consistent values (dashed);  the bottom row shows the update contributions from observations ({Term~2 in \ref{eq:cost_function_regularised}}, green) and parameter regularisation ({Term~3 in \ref{eq:cost_function_regularised}}, orange).}
  \label{fig:parameter_evolution}
\end{figure}

Next, we examine the contribution from the observations and the parameter-regularisation during the analysis steps, i.e.,  Terms 2 and 3 in \eqref{eq:cost_function_regularised}. Figure~\ref{fig:parameter_evolution} illustrates this balance for $a_1$, $\alpha_{\omega_2}$, and  $\beta^*$.   
The diffusion coefficient $\alpha_{\omega_2}$ remains essentially unchanged, indicating that the literature value is already close to the optimum.  This stability contrasts with the standard EnKF behaviour (Figure~\ref{fig:convergence}b), where $\alpha_{\omega_2}$ has large uncertainty despite contributing minimally to observation fit. Without regularisation, the parameter {is unconstrained in} regions where the cost function is relatively flat with respect to $\alpha_{\omega_2}$, whereas the regularised framework maintains the parameter near its literature-consistent value when observations provide insufficient guidance. 
By contrast, 
the Bradshaw constant $a_1$ and  the destruction coefficient  $\beta^*$ show  a stronger competition before reaching equilibrium.  
For $a_1$, regularisation {(Term~3 in \ref{eq:cost_function_regularised})} provides a positive offset of about 0.02, while the observational term {(Term~2 in \ref{eq:cost_function_regularised})} gives a nearly equal negative offset, bringing the parameter close to its literature value of 0.31. Because $a_1$ enters the eddy-viscosity formulation, only small adjustments are needed to match the data. 
For $\beta^*$, the two contributions compete significantly. Observations ({Term~2}) first decrease the parameter from 0.09, while regularisation ({Term~3}) increases it. After about eight iterations, the two terms balance and the parameter mean settles near equilibrium. \\

In summary, the proposed PR-EnKF stabilises the assimilation and keeps the parameters physically meaningful, when compared with the standard EnKF. 
Appendix~\ref{app:sigma_sensitivity} includes two additional experiments to further assess the effect of regularisation strength and confirm that the converged $\sigma=0.2$ solution is a robust optimum of the cost function~\eqref{eq:cost_function_regularised}.

\subsection{Multi-field assimilation}
\label{subsec:tke_extension}

\begin{figure}[h]
  \centerline{\includegraphics[width=0.75\textwidth]{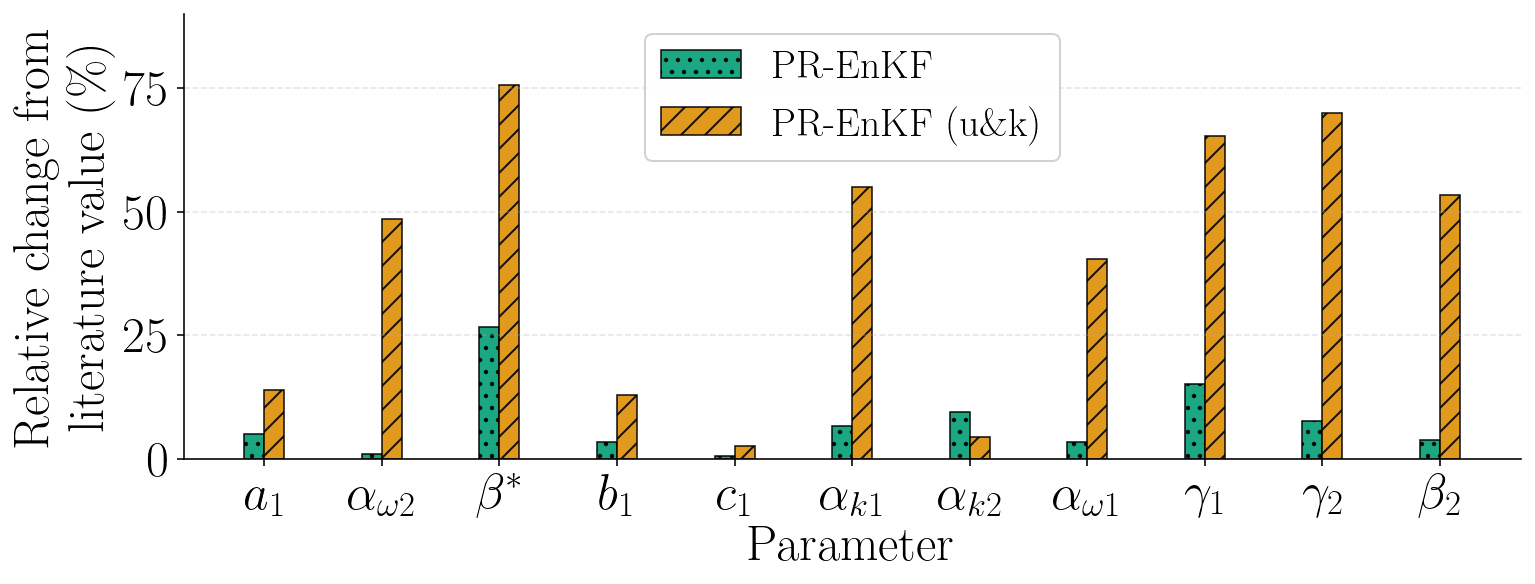}}
\caption{{Inferred} $k$-$\omega$ SST parameters from their literature-consistent values. Turbulent kinetic energy assimilation (orange, hatched) activates parameters associated with turbulence transport and dissipation, producing larger deviations than velocity-only optimisation (green, dotted).}
  \label{fig:parameter_differences}
\end{figure}
Here, we perform multi-field  assimilation (\S\ref{sec:multifield}) on the {assimilation} case (\S\ref{sec:opt_case}) and compare the results against the velocity-only {case presented in} \S~\ref{subsec:convergence_stability}.  
Table~\ref{tab:optimized_parameters} shows the multi-field {inferred} parameters in the rightmost column. Relative to the velocity-only case, the inclusion of TKE leads to marked changes in parameters that remained close to their literature-consistent values when only velocity was assimilated (highlighted rows in Table~\ref{tab:optimized_parameters}).  
The largest changes occur in parameters governing turbulence transport, production, and dissipation. In particular, $\alpha_{k1}$, $\gamma_1$, and $\gamma_2$ decrease by more than 50\%, while $\beta^*$ increases by 75\% relative to the velocity-only result. Figure~\ref{fig:parameter_differences} shows the relative difference between the default  and the optimised values with velocity-only and multi-field assimilation.   
These larger changes from the default values are consistent with the richer information content introduced by TKE. 
The regularised filter selectively updates the parameters, which differ from those updated by the EnKF.

\subsection{Prediction accuracy} 
\label{subsec:accuracy}

\begin{figure}[h]
  \centering
  \begin{subfigure}[b]{\textwidth}
    \centering
    \includegraphics[width=0.95\textwidth]{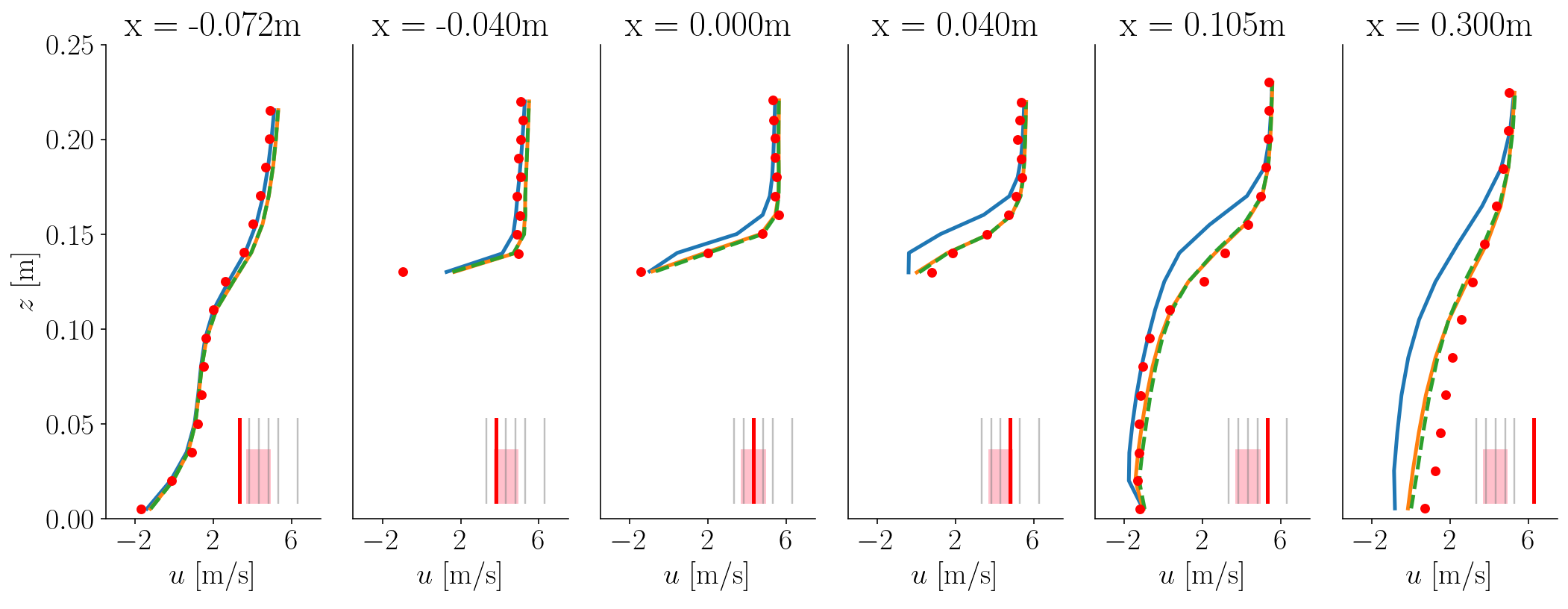}
    \caption{}
    \label{fig:profiles_a}
  \end{subfigure}


  \begin{subfigure}[b]{\textwidth}
    \centering
    \includegraphics[width=0.95\textwidth]{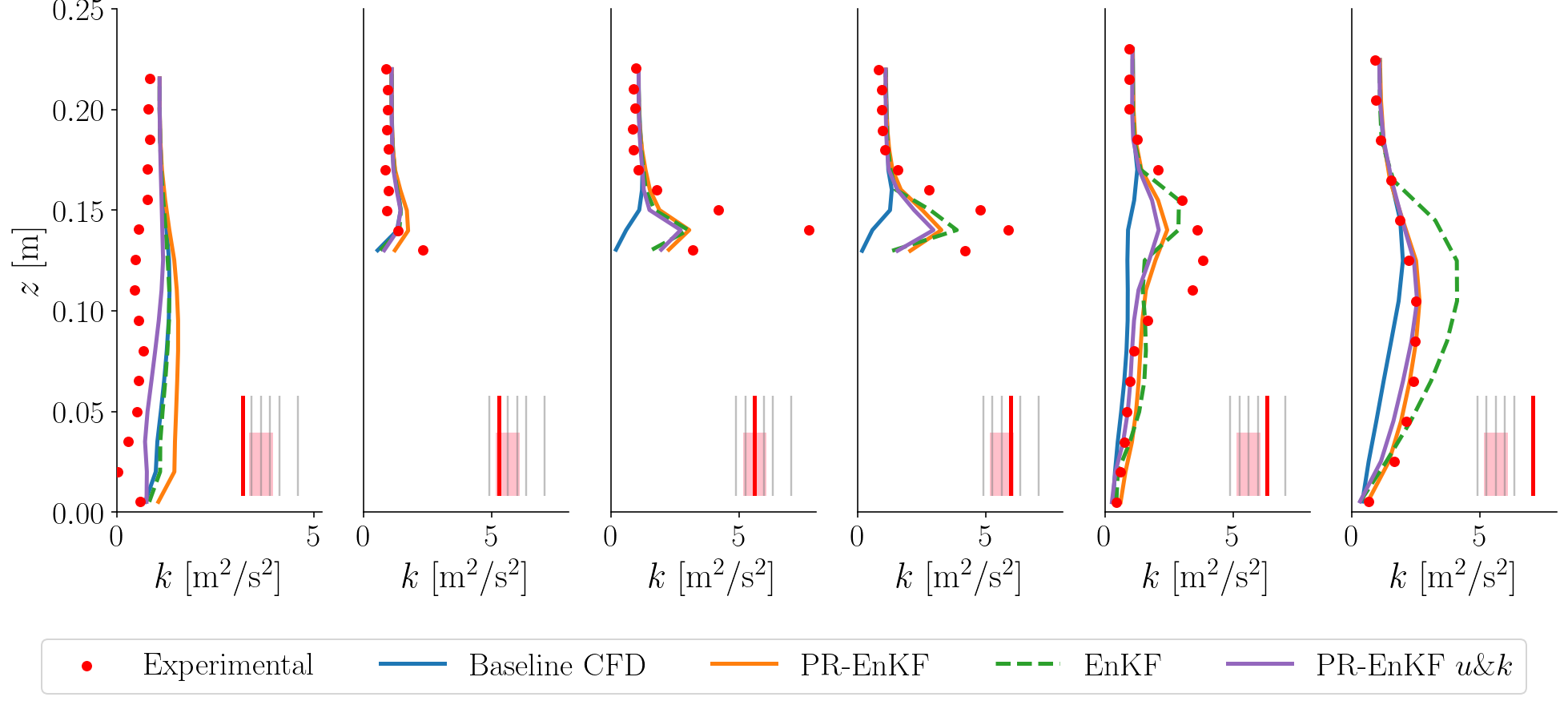}
    \caption{}
    \label{fig:profiles_b}
  \end{subfigure}

  \caption{Profiles of (a) streamwise velocity and (b) turbulent kinetic energy at six measurement locations for the CEDVAL A1-1 case \citep{CedvalHam}. Red circles show the experimental data, compared with the baseline CFD simulation (blue), EnKF (green, dashed), and the PR-EnKF with velocity-only assimilation (orange, solid) and the PR-EnKF with multi-field assimilation (purple, solid).
  The insets in each panel show the measurement positions relative to the building geometry, with the current location highlighted in red.
  The multi-field prediction is not shown in (a) for clarity, as the lines overlap with the $u$-only case.}
  \label{fig:profiles}
\end{figure}

Figure~\ref{fig:profiles} compares the experimental measurements of streamwise velocity and TKE at six locations around the CEDVAL A1-1 building (see Figure~\ref{fig1}) with the baseline CFD results and the predictions obtained using the ensemble-mean parameters from both the EnKF and PR-EnKF. 
%

Data assimilation  markedly improves the baseline CFD simulation. 
For the velocity field (Figure~\ref{fig:profiles}a), the root-mean-square error (RMSE) across the 73 measurement points decreases from $0.998\text{ m/s}$ for the baseline CFD to $0.504\text{ m/s}$ for the PR-EnKF and $0.485\text{ m/s}$ for the standard EnKF, with error reductions of $49.5\%$ and $51.4\%$, respectively. 
Although the standard EnKF does not converge, the prediction from its ensemble mean yields a marginally lower velocity RMSE in this optimisation case, which may be due to  overfitting to the observations in the CEDVAL A1-1 case.  
We evaluate whether the regularised filter enhances robustness when these parameters are transferred to unseen configurations in \S\ref{sec:results_transfer}.\\ 

The benefits of regularisation and multi-field assimilation  become apparent in the TKE profiles (Figure~\ref{fig:profiles}b). Downstream of the building ($x = 0.105$ m and $x = 0.3$ m), the standard velocity-only EnKF distorts the recirculation zone, whereas all PR-EnKF variants maintain physical consistency.  
Upstream of the building ($x = -0.072~\mathrm{m}$), incorporating TKE observations successfully suppresses the near-wall TKE overestimation produced by velocity-only assimilation. 
Quantitatively,  the EnKF achieves  a 28\% reduction in the TKE RMSE relative to the baseline CFD, whereas  the PR-EnKF assimilation yields  34\% and 38\% reductions for velocity-only and multi-field inferences, respectively. 
These results demonstrate that enriching the observation vector with TKE data constrains the turbulence field  effectively. Furthermore, they reinforce the selective activation mechanism observed in the parameter updates.

\section{Results: Transferability of the {inferred} parameters}
\label{sec:results_transfer}

The ultimate test of a turbulence parameter {inference} framework is whether it can improve the  simulation accuracy beyond the {assimilation} case. Since only velocity data are available for validation in the three transferability cases of  \S\ref{sec:transfer_cases}, we assess the transferability of the velocity-only {inferred} parameters for both the EnKF and the  PR-EnKF (with $\sigma = 0.2$), with the baseline CFD with default SST $k$-$\omega$ parameters as a reference.

\subsection{Similar Geometry -- High-rise building}
\label{Highrisebuilding}
\begin{figure}[h]
  \centerline{\includegraphics[width=0.95\textwidth]{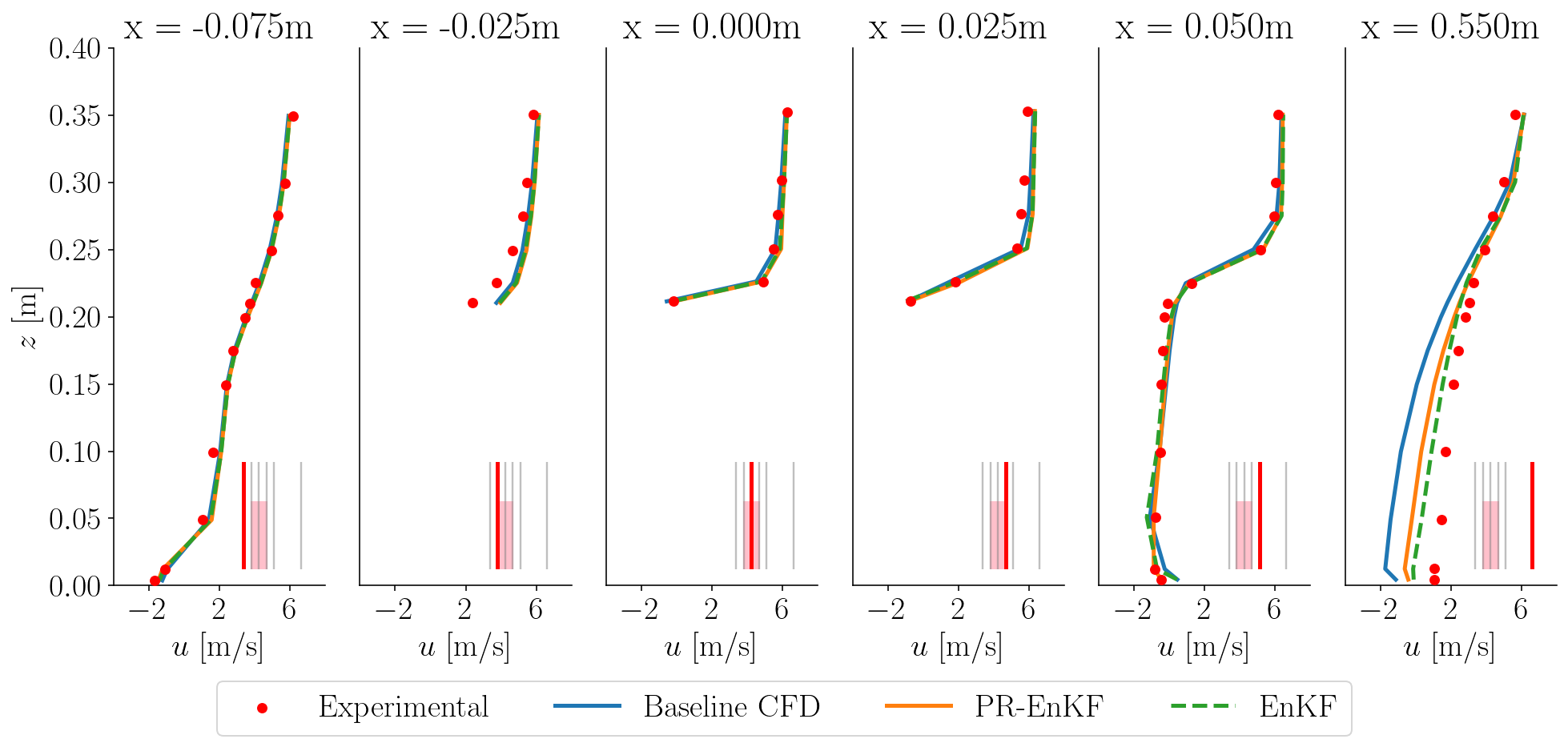}}
  \caption{Transferability to high-rise building. Streamwise velocity profiles at six measurement locations. Red circles show experimental data, compared against the baseline CFD simulation (blue, solid), PR-EnKF (orange, solid), and EnKF (green, dashed). Insets in the bottom-right indicate measurement positions relative to the building geometry, with the current location highlighted in red.}
  \label{fig:velocity_profiles_highRise}
\end{figure}

The high-rise building (Case B in \citet{AIJ_CFD_Guide_2007})  provides the most conservative transferability test, as the flow remains close to the optimisation case. Figure~\ref{fig:velocity_profiles_highRise} shows the streamwise velocity profiles at six measurement locations.  
The baseline CFD RMSE is $0.9038~\mathrm{m/s}$, the PR-EnKF reduces this to $0.6113$ {m/s}, and the standard EnKF achieves $0.5216$ m/s. 
This ordering (EnKF best, PR-EnKF intermediate, baseline worst) is consistent with the similarity between the two isolated-building configurations; the unregularised update transfers well here because the flow physics remains close to those in the optimisation case. 


\subsection{Building Array}
\label{Buildingarray}

\begin{figure}[h]
  \centering
  \includegraphics[width=0.95\textwidth]{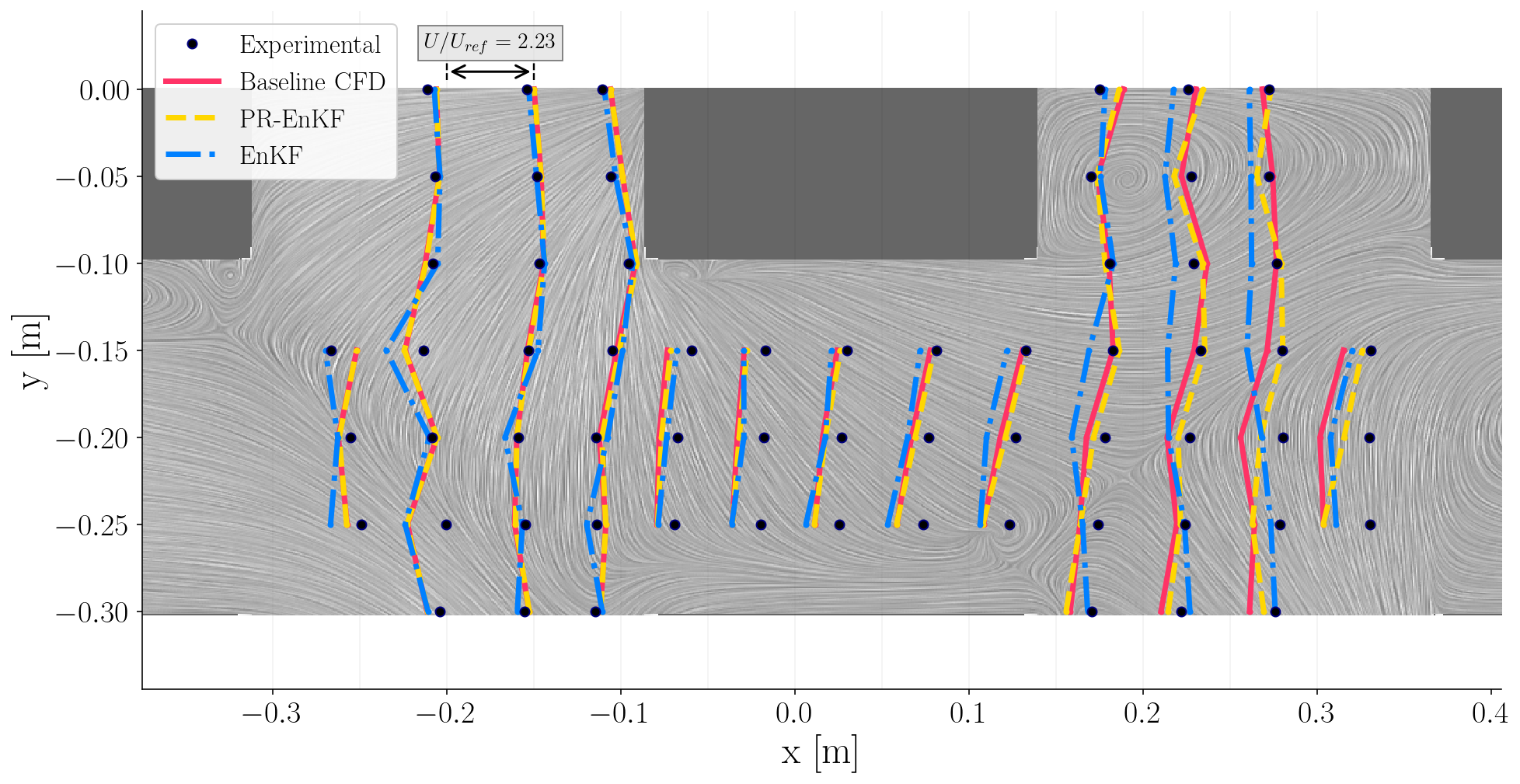}\\
  {\small (a)}\\
  \includegraphics[width=0.85\textwidth]{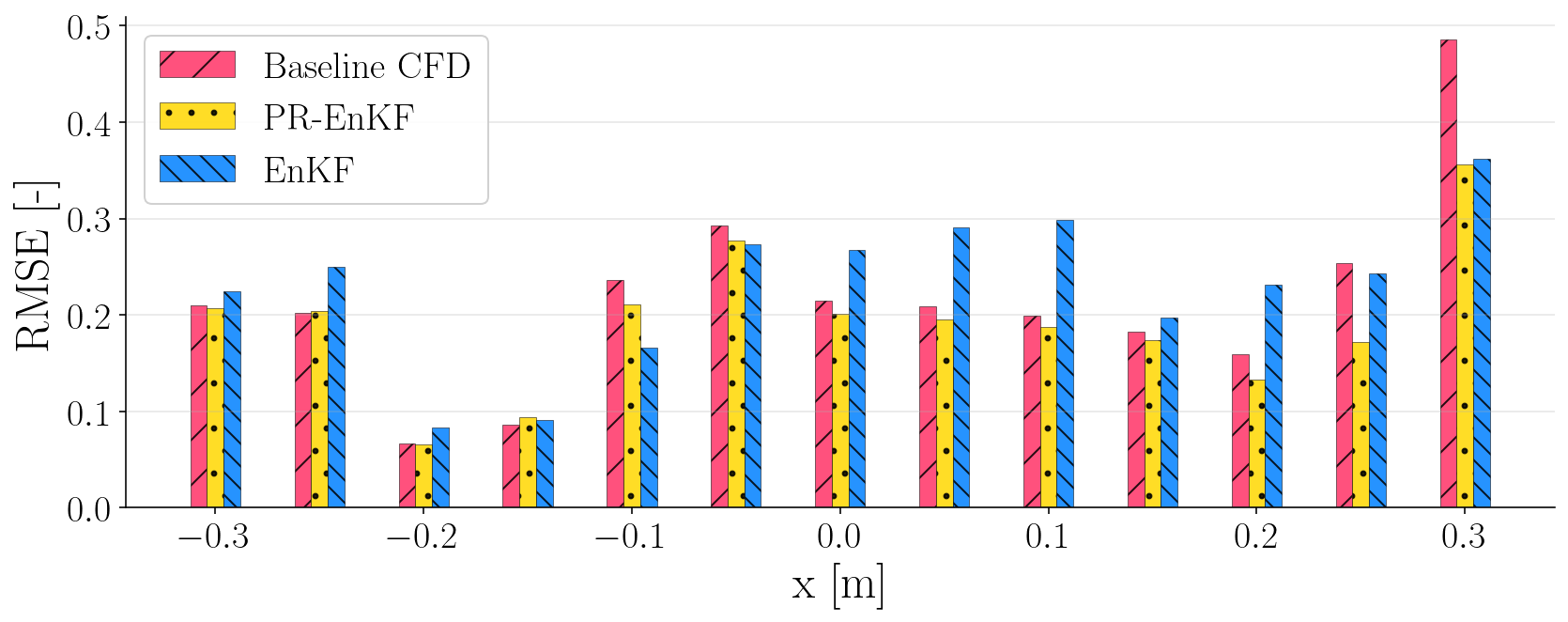}\\
  {\small (b)}
  \caption{Transferability to building array. Comparison of the predictions from the {baseline} CFD (red), the PR-EnKF (yellow), and the  standard EnKF (blue) against the experimental data. 
  (a)  Plan view of the flow field at $z = 0.02$ m with the normalised velocity magnitude shown as horizontal displacement from vertical lines at each $x$-location  (reference scale: $\Delta U/U_{\text{ref}} = 2.23$). 
  The measurement locations are shown as black circles, the dark grey areas represent building footprints, and the background streamline visualisation provides qualitative context for flow patterns. 
(b)   RMSE of normalised streamwise velocity predictions at different $x$-locations. Each bar represents the RMSE computed from all measurement points at the corresponding $x$-location.
}
  \label{fig:case_C_results}
\end{figure}
%
Figure~\ref{fig:case_C_results} shows the transferability performance of the parameters {inferred} via the EnKF and PR-EnKF on the building array (Case C IIc in \citet{AIJ_CFD_Guide_2007}). This case  introduces {interacting wakes, channelling and acceleration through the street gaps}, which are absent from the isolated obstacle of the optimisation configuration. 
Figure~\ref{fig:case_C_results}a provides a plan view at {$0.02$~m above the ground} showing the horizontal flow field with experimental measurement locations (black circles) overlaid. The velocity profiles at each measurement {location}---represented by  horizontal displacements from the vertical lines---indicate where each data assimilation  approach modifies the baseline flow field, and whether those modifications align with experimental observations. 
The spatial error distribution shows differences in how regularised and unregularised optimisation transfer to unseen geometries. In the upstream region ($x = -0.3$ to $-0.2$ m), all three approaches have similar RMSE, indicating that this incoming flow region remains largely unaffected by parameter optimisation. Critical performance differences emerge in the central building array region ($x = -0.1$ to 0.2 m), where building-building interactions dominate. Here, the PR-EnKF systematically outperforms baseline across nearly every streamwise location, demonstrating the successful transferability of learned turbulence closure corrections to wake interaction zones. The standard EnKF shows erratic behaviour with slight improvements at some locations (e.g., $x = -0.05$ m) but significant degradation at others (e.g., $x = 0.0$, $0.1$, $0.2$ m).
Figure~\ref{fig:case_C_results}b quantifies these trends in terms of RMSE at each streamwise location. Overall, averaging across all 80 measurement points, the baseline CFD yields an RMSE of $0.2139$ {m/s}, the PR-EnKF reduces this to $0.1843$ m/s, and the standard EnKF increases it to $0.2250$ m/s. The PR-EnKF-{inferred} parameters are, therefore,   transferable in the building-array case (a $13.8\%$ improvement over baseline), whereas the standard EnKF does not extrapolate well and degrades the prediction relative to the baseline case. \\

\begin{figure}[h]
  \centerline{\includegraphics[width=0.95\textwidth]{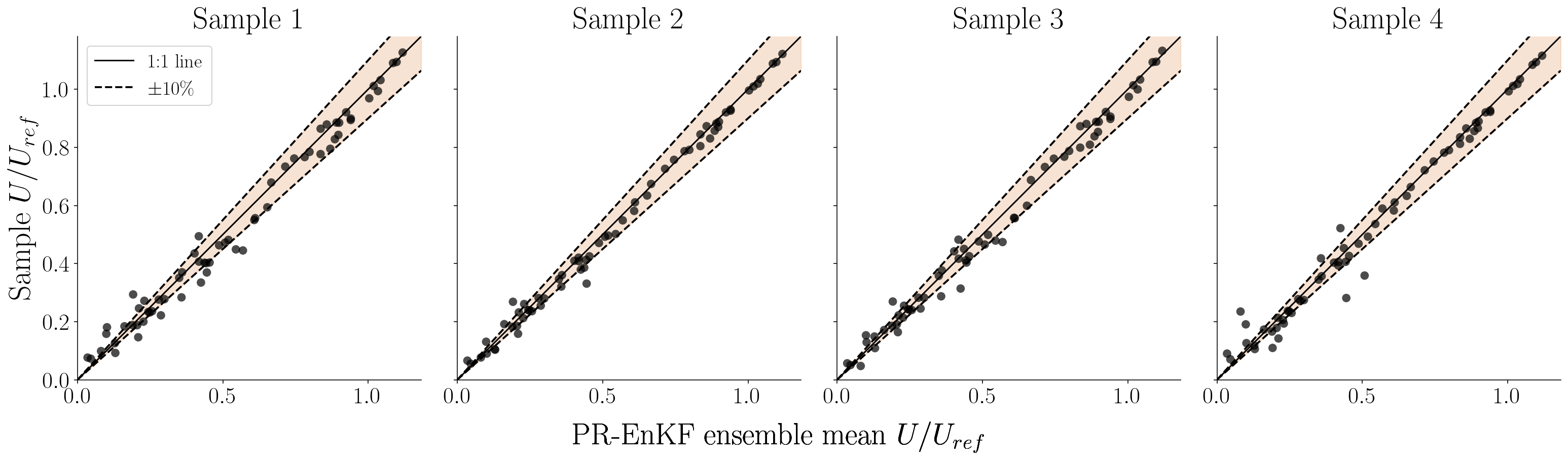}}
\caption{Transferability to building array. {Parity} plots of normalised velocity $U/U_\mathrm{ref}$ for four 
         RANS simulations using parameters  sampled from the PR-EnKF posterior, plotted against 
         the ensemble mean prediction at all 80 measurement locations.}
  \label{fig:caseC_scatter}
\end{figure}

To quantify the ensemble spread, i.e., the parameter uncertainty, we perform four additional RANS simulations using different sets of SST parameters sampled  from $\mathcal{N}(\bar{\vect{\alpha}}^a, \matr{C}^a_{\alpha\alpha})$, where $\bar{\vect{\alpha}}^a$ and $\matr{C}^a_{\alpha\alpha}$ are the ensemble mean and covariance at the final assimilation cycle.   
Table~\ref{tab:caseC_variants} shows the sampled parameters  with their prediction RMSE.   Figure~\ref{fig:caseC_scatter} shows parity plots of the four samples against the PR-EnKF ensemble mean prediction at all 80 measurement locations. All four variants yield lower RMSE than the baseline (0.2139~m/s), with values ranging from 0.1765 to 0.1918~m/s, indicating that the improvement is not limited to the ensemble mean but is generally consistent across the posterior spread. 
\begin{table}
  \begin{center}
  \def~{\hphantom{0}}
  \begin{tabular}{lccccc|c}
    \hline
    Variant & $\beta^*$ & $a_1$ & $\gamma_1$ & $\gamma_2$ & $\alpha_{k1}$ & RMSE \\
    \hline
    Ensemble mean & 0.066 & 0.294 & 0.471 & 0.406 & 0.794 & 0.1843 \\
    Sample 1    & 0.067 & 0.305 & 0.473 & 0.409 & 0.827 & 0.1765 \\
    Sample 2    & 0.069 & 0.312 & 0.489 & 0.422 & 0.848 & 0.1902 \\
    Sample 3    & 0.065 & 0.298 & 0.475 & 0.403 & 0.812 & 0.1812 \\
    Sample 4    & 0.068 & 0.308 & 0.484 & 0.418 & 0.841 & 0.1918 \\
    \hline
    Baseline     & \multicolumn{5}{c|}{default SST $k$-$\omega$} & 0.2139 \\
    \hline
  \end{tabular}
  \caption{Selected PR-EnKF posterior parameter samples and corresponding 
           velocity RMSE on AIJ Case~C.}
  \label{tab:caseC_variants}
  \end{center}
\end{table}
%


\subsection{Real-world case -- Shinjuku urban district} 
\begin{figure}[h]
  \centering
  \begin{subfigure}[b]{\textwidth}
    \centering
    \includegraphics[width=0.99\textwidth]{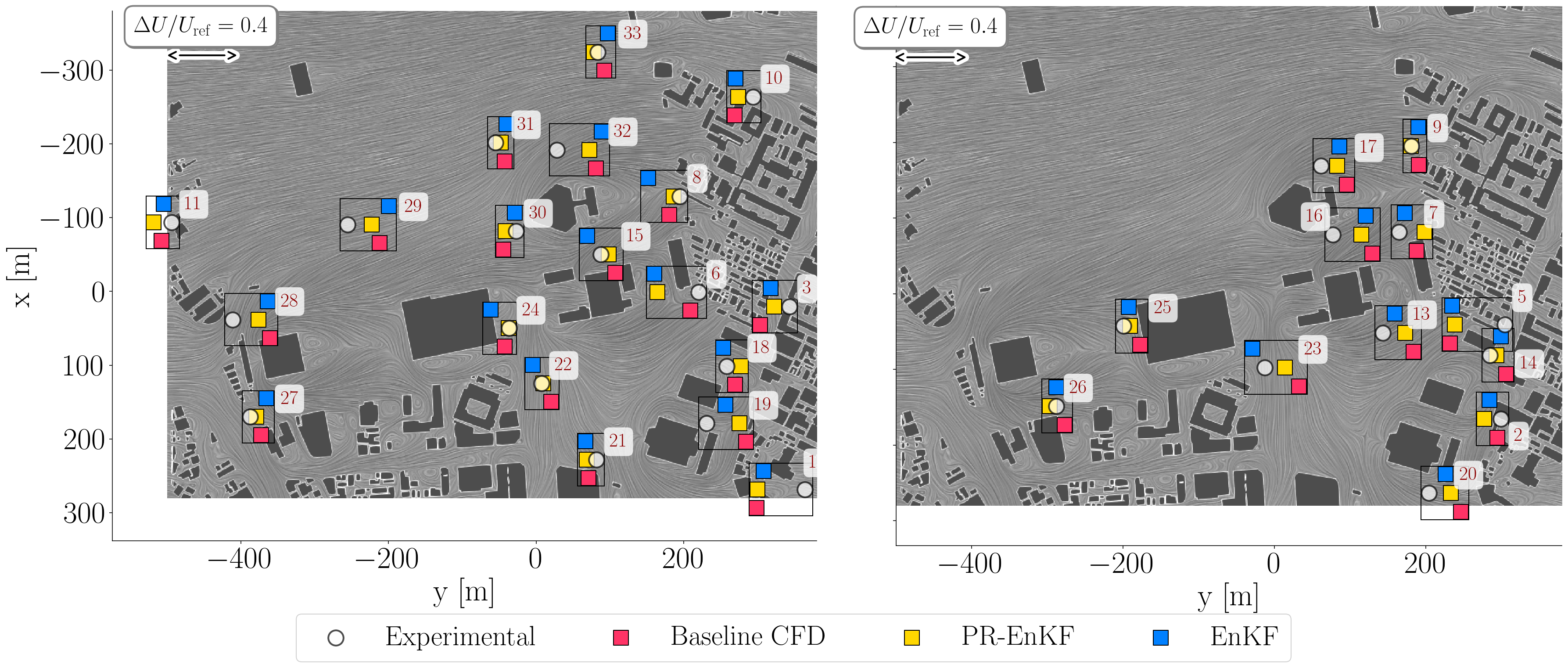}
    \caption{}
    \label{fig:shinjuku_a}
  \end{subfigure}

  \begin{subfigure}[b]{\textwidth}
    \centering
    \includegraphics[width=0.65\textwidth]{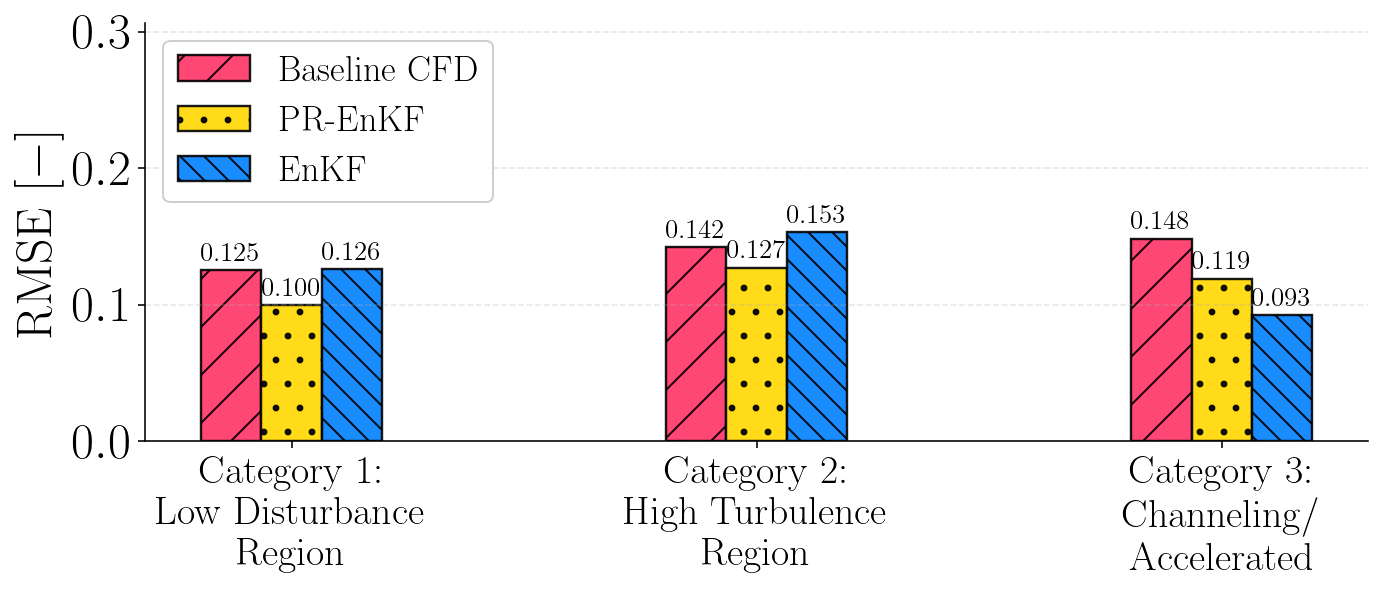}
    \caption{}
    \label{fig:shinjuku_b}
  \end{subfigure}

  \caption{Transferability to a real urban district (Shinjuku).
  Comparison of the predictions from the {baseline} CFD (red), the PR-EnKF (yellow), and the standard EnKF (blue) against the experimental data.
  (a) Plan view of the Shinjuku district showing the spatial distribution of velocity prediction errors at the 31 measurement sites ({white circles denote locations where experimental data are available}).
  Horizontal offsets indicate prediction error (reference scale: $\Delta U/U_{\text{ref}} = 0.4$). Background contours show velocity magnitude.
  (b) RMSE of normalised wind-speed ratio predictions for the measurement locations grouped into three flow categories.}
  \label{fig:shinjuku}
\end{figure}

The Shinjuku district is the most demanding transferability  test. This shows whether the parameters {inferred} on simplified laboratory configurations can improve the  predictions in a realistic urban environment. 
Figure~\ref{fig:shinjuku} shows the comparison between the RANS fields computed with the EnKF-{inferred}, PR-EnKF-{inferred} and the default SST parameters.  

The prediction errors at  the 31 measurement locations are shown by the horizontal offset in Figure~\ref{fig:shinjuku}a, and Figure~\ref{fig:shinjuku}b quantifies the prediction in terms of RMSE grouped into three flow regimes: 
Category 1 (Low Disturbance, locations 11, 26-33) represents areas with lower flow disturbance levels compared to other regions; Category 2 (High Turbulence, locations 1-3, 5, 6, 8, 10, 14, 22-25) includes complex urban regions with strong building interactions; and Category 3 (Channelling/Accelerated, locations 7, 9, 13, 15-21) includes street canyon and inter-building regions.
Across all 31 locations, the baseline CFD yields an RMSE of 0.139, the PR-EnKF reduces this to 0.116 (16.2\% improvement), and the standard EnKF reduces it to 0.125 (9.7\% improvement). The  analysis of each category shows that the PR-EnKF consistently improves the baseline solution in the three regimes and avoids the degradation of the standard EnKF in the very turbulent regions (Category 2), while the unconstrained EnKF can provide larger local gains only in strongly accelerated flows (Category 3).

\begin{figure}[h]
  \centering
  \begin{subfigure}[b]{\textwidth}
    \centering
    \includegraphics[width=0.50\textwidth]{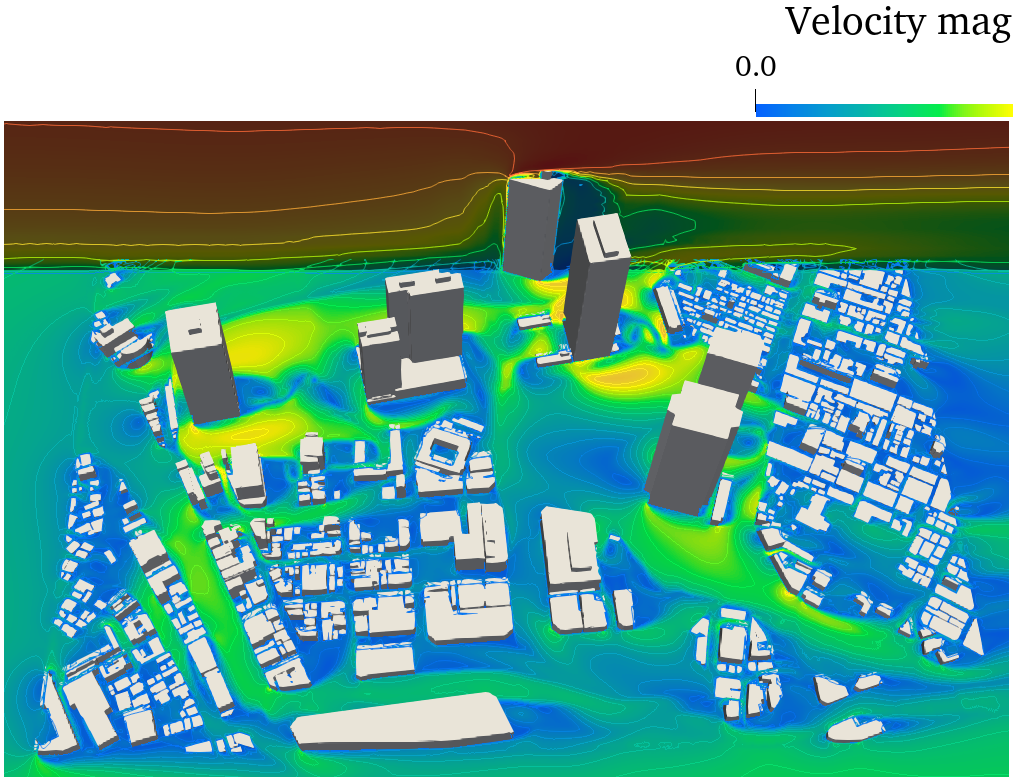}%
    \includegraphics[width=0.50\textwidth]{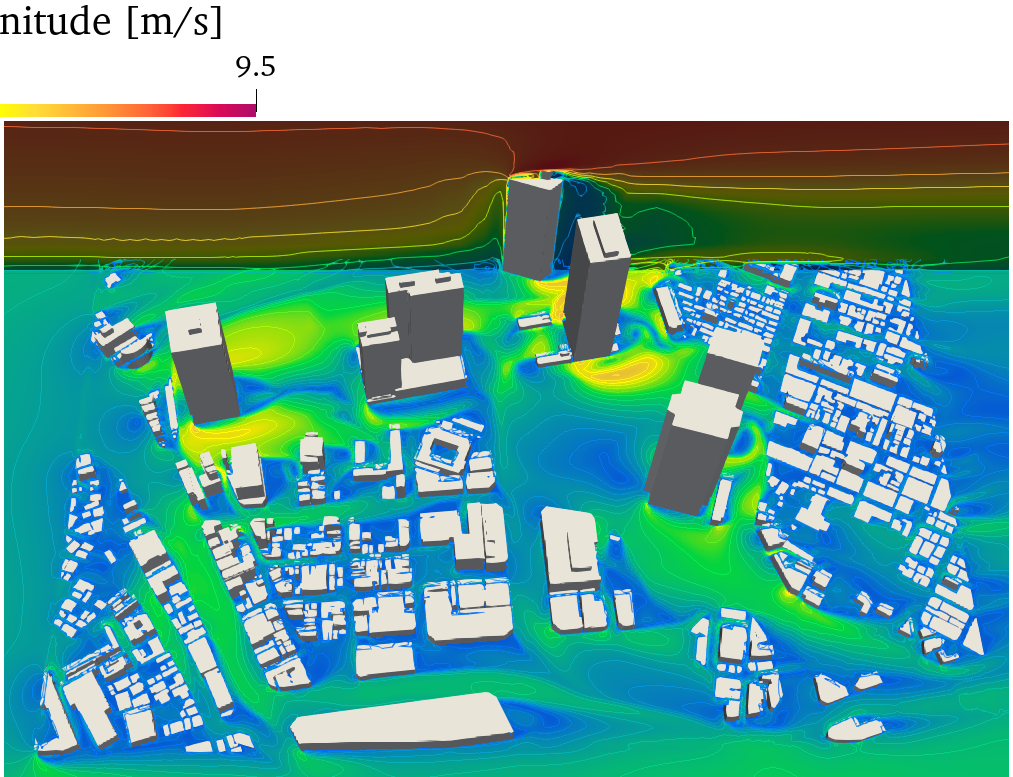}
    \caption{}
    \label{fig:shinjuku_comparison_a}
  \end{subfigure}

  \vspace{0.3cm}

  \begin{subfigure}[b]{\textwidth}
    \centering
    \includegraphics[width=0.50\textwidth]{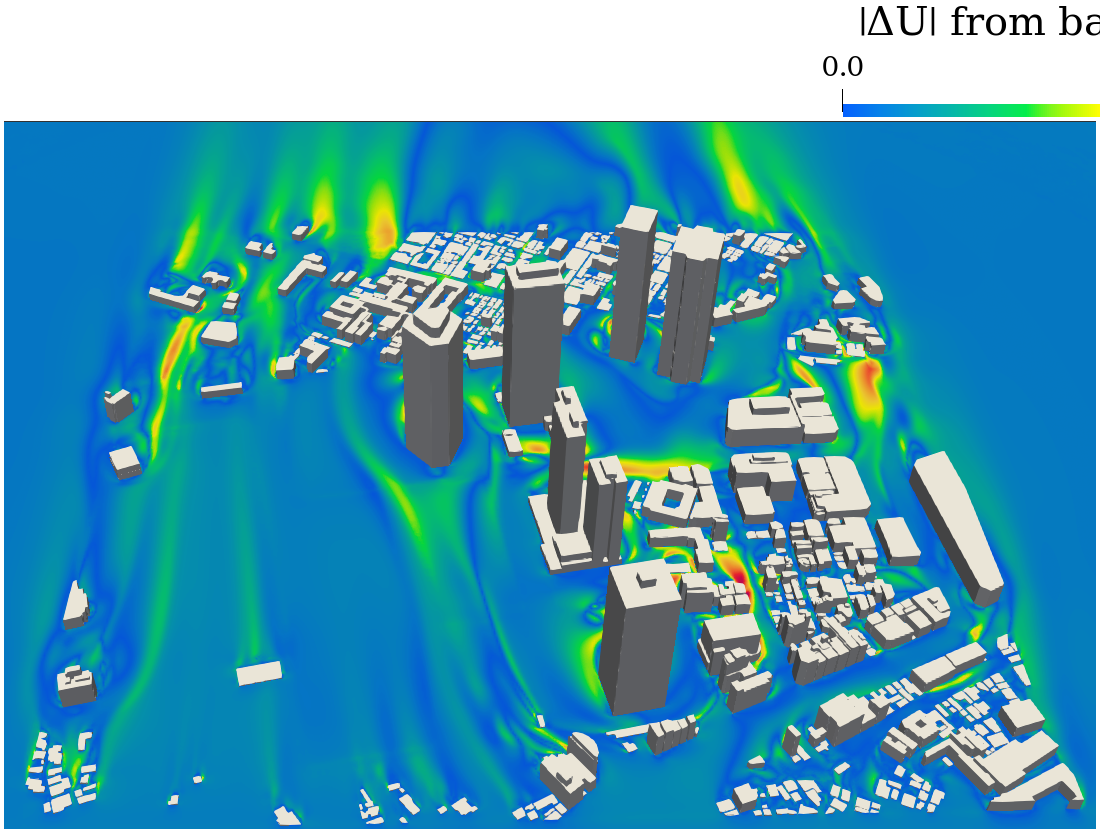}%
    \includegraphics[width=0.50\textwidth]{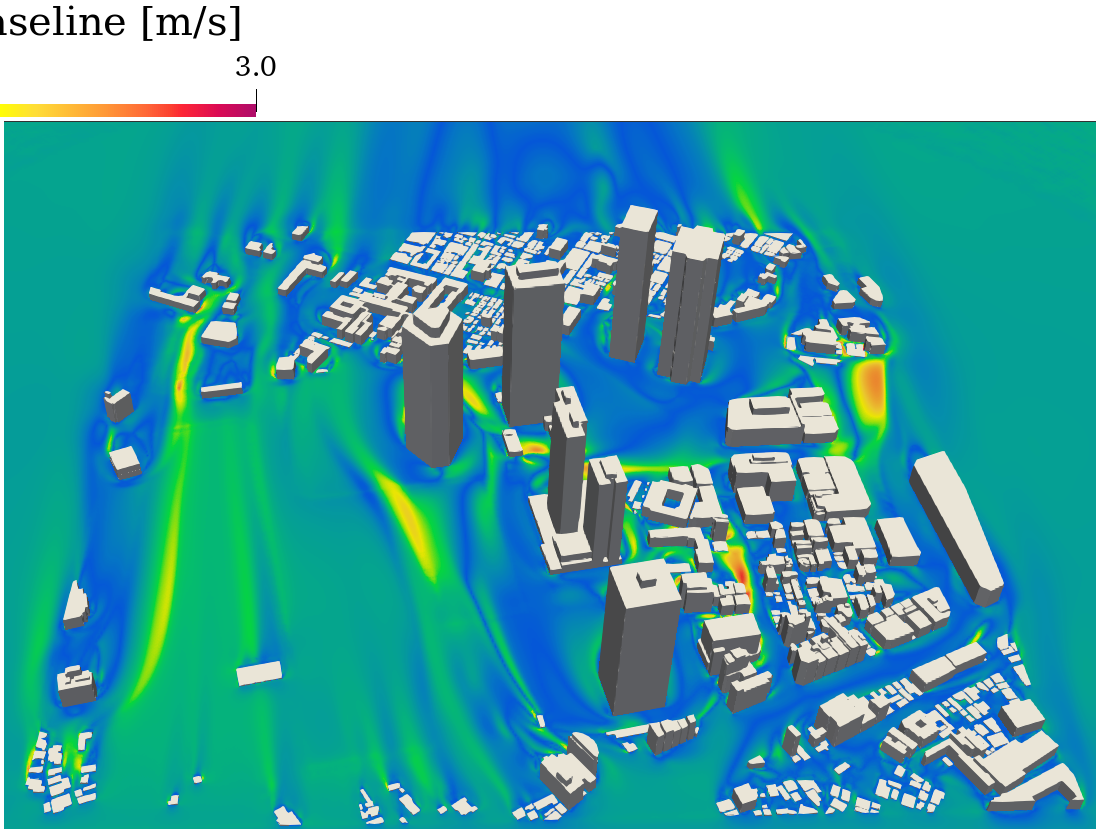}
    \caption{}
    \label{fig:shinjuku_comparison_b}
  \end{subfigure}

  \caption{Flow field comparisons in the Shinjuku urban environment obtained using the PR-EnKF (left columns) and standard EnKF (right columns). The top panels  show the velocity magnitude contours evaluated at the plane $x = -100$ m and horizontal plane $z = 10$ m.
   The bottom panels  display the absolute velocity deviation from the baseline CFD simulation ($|\Delta \boldsymbol{U}| = |\boldsymbol{U}_{\text{method}} - \boldsymbol{U}_{\text{baseline}}|$) at $z = 10~\mathrm{m}$. }
  \label{fig:shinjuku_comparison}
\end{figure}

Finally, we show the full flow field of the Shinjuku urban environment in Figure~\ref{fig:shinjuku_comparison}. 
The velocity magnitude fields in Figure~\ref{fig:shinjuku_comparison}a show the flow solution from both methods, with differences between the PR-EnKF and the standard EnKF being primarily located in the building wake regions. Figures~\ref{fig:shinjuku_comparison}b show the absolute velocity differences between each method and the baseline CFD.  The PR-EnKF produces low deviations (cool colours) in the approaching flow and low-disturbance regions, with elevated deviations concentrated in the wake zones immediately downstream of buildings---the flow regimes where observational data provided clear guidance in the optimisation case.
In contrast, the EnKF produces elevated deviations from the baseline CFD throughout the domain, including in approaching flow and low-disturbance regions. This deviation {shows} that unconstrained parameters modify the flow physics indiscriminately, lacking the selectivity to distinguish between flow regimes requiring correction, those already well-captured, and those lacking observational evidence for modification. \\

In summary, the transferability results show that RANS simulations using the PR-EnKF-optimised parameters consistently reduce the error relative to the baseline across flow regimes that are not present in the optimisation case, from the building array to the full Shinjuku district. 
In contrast, the unconstrained EnKF outperforms the PR-EnKF only when the geometry  resembles the {assimilation} case or in strongly accelerated flows; however, it does not transfer  reliably and can degrade predictions in complex urban configurations.

\section{Conclusions}
\label{sec:conclusions}
RANS simulations provide a robust framework for urban wind studies; however, they rely on closure models whose parameter transferability to realistic urban environments may be limited. This paper introduces a parameter-regularised ensemble Kalman filter (PR-EnKF) for sequential inference of RANS parameters, in which a literature-consistent regularisation constrains the Bayesian updates to physically consistent bounds. The parameters are first inferred on an isolated building, then transferred to progressively more complex cases---a high-rise building, a multi-building array, and the Shinjuku district---to assess the transferability of the PR-EnKF relative to the standard EnKF.

In the assimilation case,  all SST $k$-$\omega$ parameters inferred by the PR-EnKF converge with small uncertainty, whereas the EnKF may fail to converge in the absence of regularisation. The resulting PR-EnKF flow field reduces the RMSE against experimental data by up to 50\% relative to the baseline CFD. The transferability cases then test whether this advantage extends to unseen geometries.

Both the EnKF and PR-EnKF correctly reconstruct the high-rise building case, where the governing physical mechanisms (isolated building, Reynolds number, etc.) closely resemble those of the assimilation case. However, once multi-building wake interactions come into play, as in the multi-building array, only the PR-EnKF improves on the baseline, while the EnKF degrades it. The prior regularisation term proves key to enabling the PR-EnKF to adjust only the model parameters relevant to the assimilated data type---for instance, parameters governing turbulence transport, such as $\beta^*$, are primarily affected when assimilating turbulent kinetic energy, whereas the EnKF updates all parameters indiscriminately.

Finally, on the realistic Shinjuku district, the PR-EnKF reduces error across all flow regimes, from channelling zones to highly turbulent areas, lowering the RMSE by 16.2\% relative to the baseline, compared with 9.7\% for the standard EnKF.

This work shows that regularising the data assimilation cycle with literature-consistent prior information of the turbulence model is essential for robust inference of physical parameters in urban flow RANS simulations. The PR-EnKF offers a mathematically rigorous yet computationally tractable parameter inference method for   RANS closure calibration in large-scale urban planning applications.

\section*{Funding}
This research was funded, in whole or in part, by the European Union's Horizon Europe research and innovation programme under grant agreement No 101072559. L.M. gratefully acknowledges the Brussels Institute for Advanced Studies (BrIAS) for providing an environment that directly or indirectly supported this work. L.M. also acknowledges the ERC Starting Grant 949388.

\section*{Declaration of interests}
The authors report no conflict of interest.

\section*{Data availability statement}
The experimental data used for validation are publicly available from the CEDVAL database \citep{CedvalHam} and AIJ benchmark dataset \citep{AIJ_CFD_Guide_2007}. The computational implementations are available from the corresponding author upon request.

\appendix
\section{SST $k$-$\omega$ Turbulence Model Equations}
\label{app:sst}

The PR-EnKF optimises the 11 parameters of the SST $k$-$\omega$ turbulence model \citep{Menter1994SST}, listed in Table~\ref{tab:sst_parameters}. This appendix provides the complete model equations to clarify the physical role of each parameter.

The transport equations for turbulent kinetic energy $k$ and specific dissipation rate $\omega$ are:
\begin{equation}
\frac{\partial k}{\partial t} + (\vec{u} \cdot \nabla) k = \nabla \cdot \left[\frac{\mu_t}{\alpha_k} \nabla k\right] + G_k - \beta^* k \omega,
\label{eq:sst_k}
\end{equation}
\begin{equation}
\frac{\partial \omega}{\partial t} + (\vec{u} \cdot \nabla) \omega = \nabla \cdot \left[\frac{\mu_t}{\alpha_\omega}\nabla \omega\right] + \frac{\gamma}{\mu_t} G_k - \beta \omega^2 + D_\omega,
\label{eq:sst_omega}
\end{equation}
where $G_k = \mu_t S^2$ is the turbulent kinetic energy production with $S$ the strain rate magnitude, and $D_\omega$ is the cross-diffusion term:
\begin{equation}
D_\omega = 2\left(1-F_1\right) \alpha_{\omega_2} \frac{1}{\omega} \nabla k \cdot \nabla \omega.
\label{eq:cross_diffusion}
\end{equation}
The blended coefficients $\alpha_k$, $\alpha_\omega$, $\gamma$, and $\beta$ transition smoothly between the $k$-$\omega$ (inner, subscript 1) and $k$-$\varepsilon$ (outer, subscript 2) formulations via the blending function $F_1$:
\begin{equation}
\phi = F_1 \phi_1 + (1-F_1)\phi_2, \qquad \phi \in \{\alpha_k,\, \alpha_\omega,\, \gamma,\, \beta\},
\label{eq:blending}
\end{equation}
yielding the four parameter pairs $(\alpha_{k_1}, \alpha_{k_2})$, $(\alpha_{\omega_1}, \alpha_{\omega_2})$, $(\gamma_1, \gamma_2)$, and $(\beta_1, \beta_2)$ of Table~\ref{tab:sst_parameters}. The eddy viscosity incorporates the Bradshaw constant $a_1$ through a limiter preventing overestimation in adverse pressure gradient flows:
\begin{equation}
\mu_t = \frac{a_1 k}{\max\left(a_1 \omega,\, S F_2\right)},
\label{eq:eddy_visc}
\end{equation}
where $F_1$ and $F_2$ are blending functions transitioning between model regions based on wall distance $d$:
\begin{equation}
F_1 = \tanh \left[\left(\min \left[\max \left(\frac{\sqrt{k}}{\beta^* \omega d},\, \frac{500 \nu}{d^2 \omega}\right),\, \frac{4 \alpha_{\omega_2} k}{D_\omega d^2}\right]\right)^4\right],
\label{eq:F1}
\end{equation}
\begin{equation}
F_2 = \tanh \left[\left(\max \left(\frac{2\sqrt{k}}{\beta^* \omega d},\, \frac{500 \nu}{d^2 \omega}\right)\right)^2\right].
\label{eq:F2}
\end{equation}
The remaining parameters $b_1$ and $c_1$ appear in the production limiter, which bounds $G_k$ to prevent unphysical build-up in stagnation regions, and in the realisability constraint on eddy viscosity:
\begin{equation}
G_k \leftarrow \min\left(G_k,\, c_1 \beta^* k \omega\right),
\quad \text{and} \quad
\mu_t \leftarrow \min\left(\mu_t,\, \frac{b_1 k}{S}\right).
\label{eq:real_constraint}
\end{equation}

\section{Sensitivity to regularisation strength and warm-start {inference}}
\label{app:sigma_sensitivity}

\begin{figure}[h]
  \centerline{\includegraphics[width=\textwidth]{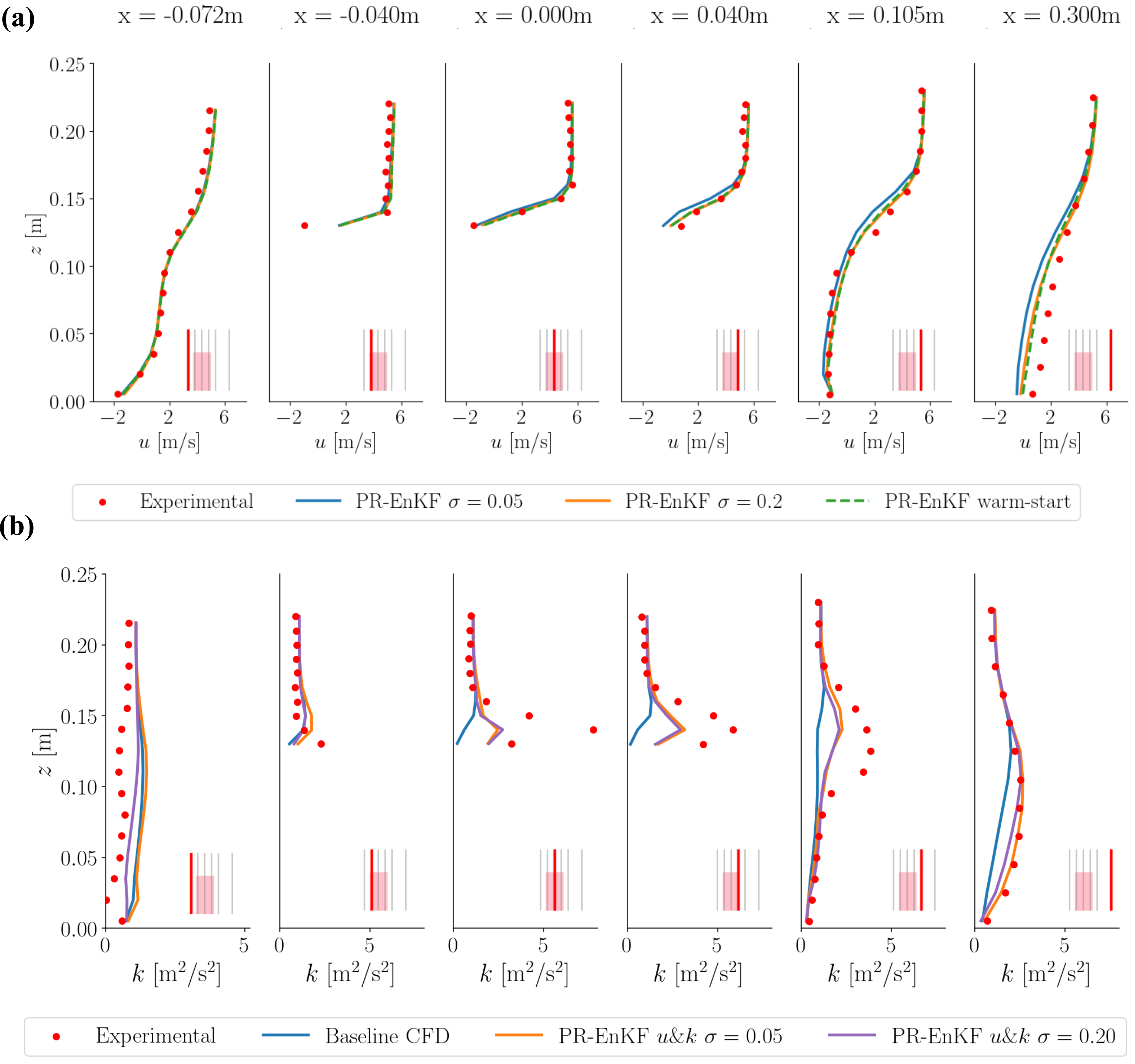}}
  \caption{Profiles of (a) streamwise velocity and (b)  turbulent kinetic energy  at six measurement locations for the CEDVAL A1-1 case \citep{CedvalHam}. 
  Red circles show the experimental data, and the insets indicate measurement positions relative to the building geometry, with the current location highlighted in red.  
    (a) Velocity-only assimilation:  comparison between the PR-EnKF with $\sigma = 0.05$ (blue, solid), $\sigma = 0.2$ (orange, solid), and the PR-EnKF warm-start (green, dashed). 
    (b) Multi-field assimilation: comparison between the baseline CFD (blue), and the PR-EnKF predictions with $\sigma = 0.05$ (orange), and $\sigma = 0.20$ (purple).
  }
  \label{fig:sigma_comparison}
\end{figure}

We test two additional CEDVAL~A1-1 runs: a stronger regularisation case with $\sigma=0.05$ and a warm-start re-{inference} initialised from the converged $\sigma=0.2$ ensemble from \S\ref{sec:opt_case}. 
Table~\ref{tab:sigma_comparison} summarises the final parameters and RMSE, and Figure~\ref{fig:sigma_comparison} compares the predicted profiles.
For the velocity-only assimilation, stronger regularisation ($\sigma = 0.05$) keeps the parameters closer to their literature values (e.g., see $\gamma_1$, and $\gamma_2$ in Tab.~\ref{tab:sigma_comparison}), ensemble standard deviations are an order of magnitude smaller than those of $\sigma = 0.2$, but this increases the velocity RMSE to 0.589~m/s. {This indicates that Term~3 in \eqref{eq:cost_function_regularised} dominates the data misfit (Term~2), over-constraining the analysis.}
The warm-start run gives a slightly lower RMSE of 0.482~m/s, with small parameter shifts but ensemble spreads narrowing considerably, which confirms that the $\sigma=0.2$ solution already lies near a cost-function minimum. 
Figure~\ref{fig:sigma_comparison}a shows the resulting velocity profiles, confirming that $\sigma = 0.2$ achieves the best compromise between data fit and parameter regularity across the three configurations. 
Figure~\ref{fig:sigma_comparison}b presents the TKE profiles obtained from the multi-field PR-EnKF with different regularisation strengths.

For the multi-field case, stronger regularisation constrains parameters more tightly to their literature-consistent values, as shown in Table~\ref{tab:sigma_comparison}. The selective activation mechanism driven by TKE cross-covariances persists across regularisation strengths, though its magnitude is modulated by $\sigma$. The resulting improvement in $k$ RMSE is approximately 35\%, lower than the 38\% achieved at $\sigma = 0.20$, consistent with the velocity-only sensitivity results.

\begin{table}
  \begin{center}
  \def~{\hphantom{0}}
\begin{small}
  \begin{tabular}{lcccccc}
   &  &  \multicolumn{5}{c}{PR-EnKF} \\ \cline{3-7}
  \multicolumn{1}{c}{Parameter} & Default & 
  $\begin{array}{c} \sigma=0.05 \\ u\text{-only} \end{array}$ & 
  $\begin{array}{c} \sigma=0.2 \\ u\text{-only} \end{array}$ & 
  $\begin{array}{c} \text{warm-start} \\ u\text{-only} \end{array}$ & 
  $\begin{array}{c} \sigma=0.05 \\ u\&k \end{array}$ & 
  $\begin{array}{c} \sigma=0.2 \\ u\&k \end{array}$ \\ \hline
  $a_1$              & 0.31   & $0.323 \pm 0.003$ & $0.294 \pm 0.026$ & $0.297 \pm 0.010$ & $0.346 \pm 0.008$ & $0.353 \pm 0.031$ \\
  $\alpha_{\omega2}$ & 0.856  & $0.863 \pm 0.010$ & $0.865 \pm 0.099$ & $0.899 \pm 0.038$ & $0.688 \pm 0.005$ & $0.440 \pm 0.019$ \\
  $\beta^*$          & 0.09   & $0.076 \pm 0.001$ & $0.066 \pm 0.009$ & $0.062 \pm 0.002$ & $0.112 \pm 0.004$ & $0.158 \pm 0.016$ \\
  $b_1$              & 1.0    & $0.963 \pm 0.011$ & $1.035 \pm 0.092$ & $1.065 \pm 0.035$ & $0.950 \pm 0.024$ & $1.130 \pm 0.097$ \\
  $c_1$              & 10.0   & $9.976 \pm 0.141$ & $10.066 \pm 1.170$ & $9.730 \pm 0.518$ & $10.700 \pm 0.318$ & $10.256 \pm 1.223$ \\
  $\alpha_{k1}$      & 0.85   & $0.828 \pm 0.009$ & $0.794 \pm 0.103$ & $0.734 \pm 0.037$ & $0.403 \pm 0.022$ & $0.382 \pm 0.090$ \\
  $\alpha_{k2}$      & 1.0    & $1.025 \pm 0.009$ & $0.905 \pm 0.115$ & $0.864 \pm 0.039$ & $0.950 \pm 0.052$ & $0.956 \pm 0.203$ \\
  $\alpha_{\omega1}$ & 0.50   & $0.501 \pm 0.005$ & $0.517 \pm 0.060$ & $0.567 \pm 0.024$ & $0.680 \pm 0.007$ & $0.298 \pm 0.027$ \\
  $\gamma_1$         & 0.555  & $0.544 \pm 0.006$ & $0.471 \pm 0.061$ & $0.421 \pm 0.026$ & $0.417 \pm 0.007$ & $0.193 \pm 0.029$ \\
  $\gamma_2$         & 0.44   & $0.435 \pm 0.006$ & $0.406 \pm 0.055$ & $0.353 \pm 0.017$ & $0.298 \pm 0.012$ & $0.132 \pm 0.049$ \\
  $\beta_2$          & 0.0828 & $0.085 \pm 0.001$ & $0.086 \pm 0.010$ & $0.105 \pm 0.004$ & $0.112 \pm 0.004$ & $0.127 \pm 0.015$ \\ \hline
  $u$-RMSE (m/s)        & 0.998  & 0.589 & 0.504 & 0.482 & 0.612 & 0.516 \\
  $k$-RMSE (m$^2$/s$^2$) & 1.421  & 1.035 & 0.938 & 0.935 & 0.923 & 0.881 \\ \hline
  \end{tabular}
  \caption{SST $k$-$\omega$ parameters and RMSE after 40 assimilation cycles for different PR-EnKF regularisation configurations on the CEDVAL~A1-1 case. Default values are from \citet{Menter1994SST}. }
  \label{tab:sigma_comparison}
  
\end{small}
  \end{center}
\end{table}


%



\bibliographystyle{unsrtnat}

\bibliography{biblio.bib}




\end{document}